\pdfoutput=1
\documentclass[onecolumn,12pt,twoside]{IEEEtran}
\ifCLASSINFOpdf
   \usepackage[pdftex]{graphicx}
   \graphicspath{{../pdf/}{../jpeg/}}
   \DeclareGraphicsExtensions{.pdf,.jpeg,.png}
\else
   \usepackage[dvips]{graphicx}
   \graphicspath{{../eps/}}
   \DeclareGraphicsExtensions{.eps}
\fi
\usepackage[caption=false,font=normalsize,labelfont=sf,textfont=sf]{subfig}
\usepackage[cmex10]{amsmath}
\usepackage{amsfonts}
\usepackage{amssymb}
\usepackage{amsthm}
\usepackage{cite}
\usepackage{bm}
\usepackage{setspace}
\usepackage[caption=false,font=normalsize,labelfont=sf,textfont=sf]{subfig}

\begin{document}
%
\bibliographystyle{IEEEtran}
\title{A Comparative Study of Downlink MIMO Cellular Networks with Co-located and Distributed Base-Station Antennas}
\author{Zhiyang~Liu,~\IEEEmembership{Student~Member,~IEEE,} and~Lin~Dai,~\IEEEmembership{Senior~Member,~IEEE}

\thanks{Z. Liu and L. Dai are with the Department
of Electronic Engineering, City University of Hong Kong,
83 Tat Chee Avenue, Kowloon Tong, Hong Kong, China (email: lzy.ee@my.cityu.edu.hk;
lindai@cityu.edu.hk). }}
\maketitle


\begin{abstract}

Despite the common belief that substantial capacity gains can be achieved by using more antennas at the base-station (BS) side in cellular networks, the effect of BS antenna topology on the capacity scaling behavior is little understood. 
In this paper, we present a comparative study on the ergodic capacity of a downlink single-user multiple-input-multiple-output (MIMO) system where BS antennas are either co-located at the center or grouped into uniformly distributed antenna clusters in a circular cell. By assuming that the number of BS antennas and the number of user antennas go to infinity with a fixed ratio $L\gg 1$, the asymptotic analysis reveals that the average per-antenna capacities in both cases logarithmically increase with $L$, but in the orders of $\log_2 L$ and $\tfrac{\alpha}{2}\log_2 L$, for the co-located and distributed BS antenna layouts, respectively, where $\alpha>2$ denotes the path-loss factor. The analysis is further extended to the multi-user case where a 1-tier (7-cell) MIMO cellular network with $K\gg 1$ uniformly distributed users in each cell is considered. By assuming that the number of BS antennas and the number of user antennas go to infinity with a fixed ratio $L\gg K$, an asymptotic analysis is presented on the downlink rate performance with block diagonalization (BD) adopted at each BS. It is shown that the average per-antenna rates with the co-located and distributed BS antenna layouts scale in the orders of $\log_2 \tfrac{L}{K}$ and $\log_2 \frac{(L-K+1)^{\alpha/2}}{K}$, respectively.
The rate performance of MIMO cellular networks with small cells is also discussed, which highlights the importance of employing a large number of distributed BS antennas for the next-generation cellular networks.

\end{abstract}

\newpage

\section{Introduction}

The next-generation cellular networks are expected to provide high data rates to support the massive mobile applications. Towards this end, there has been a growing interest in implementing large antenna arrays at the base stations (BSs)\cite{Marzetta2010,Zakhour2012,Rusek2013,JSAC_Large_Scale}. It is well-known that for a point-to-point multiple-input-multiple-output (MIMO) system with $M$ transmit and $N$ receive antennas, the capacity grows linearly with $\min(M,N)$ in a rich-scattering environment \cite{Telatar1999}. With a large number of co-located antennas at both the BS and the user sides, nevertheless, the capacity may be severely reduced due to strong antenna correlation \cite{Chizhik2002}.

If the BS antennas are grouped into geographically distributed clusters and connected to a central processor by fiber or coaxial cable, in contrast, signals from distributed BS antennas to each user are subject to independent and different levels of large-scale fading, thanks to which potential capacity gains over the co-located counterpart can be expected\cite{Roh2002,Zhuang2003,Zhang2004}. In the meanwhile, the implementation cost of distributed BS antennas also becomes significantly higher than that of the co-located ones, especially when the number of distributed BS antenna clusters is large. It is, therefore, of great practical importance to compare the rate performance of cellular networks under different BS antenna layouts to see if the increased cost is justified. In this paper, we will present a comparative study on the downlink rate performance of MIMO cellular networks with co-located and distributed BS antennas, and explore how the rate scaling behavior varies with different BS antenna layouts when a large number of BS antennas are employed.

\subsection{Single-User Capacity}

In the single-user case, the ergodic capacity of a point-to-point MIMO channel has been extensively studied in the past decade. With co-located antennas at both sides, all the transmit signals experience the same large-scale fading, and thus the ergodic capacity can be fully described as a function of the average received signal-to-noise ratio (SNR)\cite{Telatar1999}. Asymptotic results from random matrix theory \cite{Tulino2004,Couillet2011} were also successfully applied to characterize the ergodic capacity when the number of antennas is large \cite{Lozano2002,Tulino2005}. By assuming that the number of antennas on both sides grow infinitely with a fixed ratio, the asymptotic ergodic capacity of a point-to-point MIMO channel was shown to be solely determined by the average received SNR and the ratio of the number of transmit antennas to the number of receive antennas \cite{Lozano2002}.

With distributed BS antennas, in contrast, the ergodic capacity is further determined by the positions of the user and BS antennas \cite{Aktas2006,Feng2009b,Heliot2011,Zhang2013b,Zhuang2003,Choi2007,Lee2012, Wang2013}. By assuming that BS antennas are grouped into $L$ geographically distributed antenna clusters, and the number of antennas at each cluster and the number of user antennas grow infinitely with a fixed ratio, the asymptotic ergodic capacity of a distributed MIMO channel was derived in \cite{Feng2009b,Aktas2006,Heliot2011,Zhang2013b} as an implicit function of $L$ large-scale fading coefficients. As the positions of BS antennas and the user may vary under different scenarios, the \textit{average ergodic capacity} was considered in \cite{Zhuang2003,Choi2007,Lee2012,Wang2013}, where the ergodic capacity is averaged over the large-scale fading coefficients from distributed BS antenna clusters to the user.
When the number of BS antenna clusters $L$ is large, nevertheless, it becomes increasingly difficult to obtain the average ergodic capacity due to high computational complexity. How the average ergodic capacity scales with $L$ has thus remained largely unknown.
As we will show in this paper, asymptotic bounds would be helpful for us to characterize the scaling behavior of the average ergodic capacity of distributed MIMO channels.

Specifically, we consider a downlink single-user system with $M$ BS antennas and $N$ co-located antennas at the user. Two BS antenna layouts are considered: 1) the co-located antenna (CA) layout where the BS antennas are co-located at the center of the cell, and 2) the distributed antenna (DA) layout where the BS antennas are grouped into $\tfrac{M}{N}$ clusters which are uniformly distributed within the inscribed circle of the hexagonal cell. In contrast to most previous studies where a regular BS antenna layout is adopted\cite{Choi2007,Lee2012,Heliot2011,Aktas2006,Wang2013,Zhang2013b}, we assume a random BS antenna layout because 1) when the number of BS antenna clusters is large, it is difficult to place them in a regular manner due to complicated geographic conditions, and 2) a random BS antenna layout describes a more general scenario and provides a reasonable performance lower-bound.

Note that the channel state information (CSI) was usually assumed to be absent at the transmitter side in previous studies \cite{Roh2002,Zhuang2003,Zhang2004,Choi2007,Feng2009b,Lee2012,Aktas2006, Heliot2011,Zhang2013b,Wang2013}. With $M \gg N$, i.e., much more transmit antennas than receive antennas, substantial capacity gains can be achieved by optimally allocating the transmit power according to CSI. It is, therefore, of great importance to study the capacity with CSI at the transmitter side (CSIT) of the distributed MIMO channel. In this paper, we assume that perfect CSI is available at both the BS and the user sides, and present an asymptotic analysis of the per-antenna capacity with $M,N\to\infty$ and $M/N\to L\gg1$. The asymptotic per-antenna capacity with the CA layout and an asymptotic lower-bound of the per-antenna capacity with the DA layout are derived, both of which are found to be closely dependent on the minimum access distance of the user. The average per-antenna capacity, which is obtained by averaging over the large-scale fading coefficients, is further analyzed in both cases. The analysis shows that the asymptotic average per-antenna capacity with the CA layout and the asymptotic lower-bound of the average per-antenna capacity with the DA layout both logarithmically increase with $L$, but in the orders of $\log_2L$ and $\tfrac{\alpha}{2}\log_2 L$, respectively, where $\alpha>2$ denotes the path-loss factor.\footnote{Note that for metropolitan areas where the propagation loss is high, the path-loss factor $\alpha$ could be much larger than $2$, in which case the average per-antenna capacity with the DA layout increases with the ratio $L$ of the number of BS antennas to the number of user antennas at a significantly higher rate than that with the CA layout.} When the ratio $L$ of the number of BS antennas to the number of user antennas is large, a much higher capacity is  achieved in the DA case thanks to the reduction of the minimum access distance.

\subsection{Multi-User Rate}

In a multi-user cellular system, the downlink rate performance of each user is crucially determined by the precoding strategy. Various precoding schemes have been proposed (see \cite{Gesbert2007} for a comprehensive overview), among which an orthogonal linear precoding scheme, block diagonalization (BD)\cite{Spencer2004}, has gained widespread popularity thanks to its low complexity and near-capacity performance when the number of BS antennas is large\cite{Shen2006,Shen2007,Shim2008,Ravindran2008}.

With BD, the intra-cell interference is eliminated by projecting the user's signal to the null space of all other users' channel gain matrices. With co-located BS antennas in each cell, the asymptotic per-user rate of a downlink cellular system with BD was recently characterized in \cite{Liu2012} by assuming that the number of BS antennas and the number of user antennas go to infinity with a fixed ratio. It was shown that with equal power allocation among users, the asymptotic rate is sensitive to the user's position, and logarithmically increases with the ratio of the number of BS antennas to the number of user antennas. If the BS antennas are geographically distributed, the rate performance is further dependent on the BS antennas' positions. For computational tractability, most studies have focused on a regular BS antenna layout with a small number of BS antennas\cite{Li2009,Wang2009a,Ahmad2011,Heath2011}.
In this paper, an asymptotic lower-bound will be developed to characterize the scaling behavior of the average rate performance with BD when the number of BS antenna clusters and the number of users are large.

Specifically, we consider a 1-tier (7-cell) cellular system with $K\gg 1$ uniformly distributed users each equipped with $N$ co-located antennas in each cell, and $M$ BS antennas either co-located at the center of each cell or grouped into $\frac{M}{N}$ uniformly distributed clusters. By assuming $M,N\to\infty$ and $M/N\to L\gg K$, an asymptotic lower-bound of the average per-antenna rate with BD in the DA layout is derived, and compared with the asymptotic average rate in the CA layout. It is shown that in contrast to the CA case where the asymptotic average per-antenna rate increases in the order of $\log_2\frac{L}{K}$, the asymptotic lower-bound of the average per-antenna rate with the DA layout has a larger scaling order of $\log_2 \frac{(L-K+1)^{\alpha/2}}{K}$, where $\alpha>2$ is the path-loss factor. Simulation results verify that the average per-antenna rate in the DA layout has the same scaling order as its asymptotic lower-bound, and is much higher than that with the CA layout when the ratio $L$ of the number of BS antennas to the number of user antennas is large.

Despite substantial gains on the average rate performance, the analysis reveals that the moments of the normalized inter-cell interference power in the DA layout are divergent at the cell edge, indicating that the rate performance becomes extremely sensitive to the user's position. Intuitively, with a large number of uniformly distributed BS antenna clusters in each cell, the chance that a cell-edge user is close to some BS antenna in the neighboring cell is significantly higher than that in the CA case. Simulation results corroborate that although the rate performance can be greatly improved on average, the rate difference among cell-edge users becomes enlarged in the DA layout.

The remainder of this paper is organized as follows. Section II introduces the system model. The asymptotic capacity analysis of the single-user case is presented in Section III, and the asymptotic average rate with BD of multi-user cellular networks is characterized in Section IV. Implications of the analysis for the cellular network design are presented in Section V, and Section VI concludes this paper.

Throughout this paper, italic letters denote scalars, and boldface upper-case and lower-case letters denote matrices and vectors, respectively. The superscripts $T$ and $\dag$ denote transpose and conjugate transpose, respectively. $\mathbb{E}[\cdot]$ denotes the expectation operator. $\|\mathbf{x}\|$ denotes the Euclidean norm of vector $\mathbf{x}$. $\textmd{Tr}\{\mathbf{X}\}$ and $\det\{\mathbf{X}\}$ denote the trace and determinant of matrix $\mathbf{X}$, respectively. $\textrm{diag}(a_1,\dots,a_N)$ denotes an $N\times N$ diagonal matrix with diagonal entries $\{a_i\}$. $\mathbf{I}_N$ denotes an $N\times N$ identity matrix. $\mathbf{0}_{N\times M}$ and $\mathbf{1}_{N\times M}$ denote $N\times M$ matrices with all entries zero and one, respectively. $\mathcal{W}_p(t,\mathbf{Q})$ denotes a $p\times p$ Wishart matrix with degrees of freedom $t$ and covariance $\mathbf{Q}$.  $|\mathcal{X}|$ denotes the cardinality of set $\mathcal{X}$.

\section{System Model}

Consider a 1-tier hexagonal cellular network with a total number of $7$ cells that share the same frequency band. Each cell has a set of users, denoted by $\mathcal{K}_i$, and a set of base-station (BS) antennas, denoted by $\mathcal{B}_i$, with $\left|{{\mathcal K}}_{i} \right|{=}K$ and $\left|{ {\mathcal B}}_{i} \right|{=}M$, $i{=}0,\dots,6$. Suppose that each user is equipped with $N\ll M$ antennas. Without loss of generality, the radius of the inscribed circle of each hexagonal cell is normalized to be 1.

Let us focus on the downlink transmission of the central cell, i.e., Cell 0. Specifically, the received signal of user $k\in \mathcal{K}_0$ can be written as
\begin{equation}\label{general system model}
{{\mathbf{y}}_k} = \underbrace{{{\mathbf{G}}_{k,{\mathcal{B}_0}}}{{\mathbf{x}}_k}}_{\text{Desired Signal}} {+}
\underbrace{{{\mathbf{G}}_{k,{\mathcal{B}_0}}}\sum\limits_{j \ne k,j \in {\mathcal{K}_0}}
{{{\mathbf{x}}_j}}}_{\text{Intra-cell Interference}}   {+}
\underbrace{\sum\limits_{i = 1}^6 {{{\mathbf{G}}_{k,{\mathcal{B}_i}}}\sum\limits_{j \in {\mathcal{K}_i}}
{{\mathbf{x}_j}} }}_{\text{Inter-cell Interference}}  {+} {{\mathbf{z}}_k},
\end{equation}
where $\mathbf{x}_j \in\mathbb{C}^{M \times 1}$ is the transmitted signal vector from BS $i$ to user $j\in\mathcal{K}_i$,
$i{=}0,\dots,6$. ${{\mathbf{z}}_k}\in\mathbb{C}^{N \times 1}$ is the additive white Gaussian noise (AWGN) at user $k$,
which has independent and identically distributed (i.i.d.) complex Gaussian entries with zero mean and variance $N_0$.
${{\mathbf{G}}_{k,{\mathcal{B}_i}}}\in\mathbb{C}^{N \times M}$ denotes the channel gain matrix between BS $i$ and user
$k$, $i{=}0,\dots,6$, which is given by
\begin{equation}\label{define of G}
{{\mathbf{G}}_{k,{\mathcal{B}_i}}}=\mathbf{\Gamma}_{k,{\mathcal{B}_i}}\circ \mathbf{H}_{k,{\mathcal{B}_i}},
\end{equation}
where $\circ$ denotes the Hadamard product. $\mathbf{H}_{k,{\mathcal{B}_i}}\in\mathbb{C}^{N\times M}$ denotes the small-scale fading matrix between
BS $i$ and user $k$ with entries modeled as i.i.d. complex Gaussian random variables with zero mean and unit
variance. ${\mathbf{\Gamma}}_{k,{\mathcal{B}_i}}\in\mathbb{C}^{N\times M}$ denotes the corresponding large-scale fading matrix, which is composed of $N$ identical row vectors $\bm{\gamma}_{k,\mathcal{B}_i}$.

We assume that each BS has full channel state information (CSI) of all users in its own cell, and no cooperation is adopted among BSs. Moreover, each user $j\in \mathcal K_i$ has full CSI of the channel from its BS to itself. With linear precoding, the transmitted signal vector for user $j$ can be written as
\begin{equation}\label{TransVec}
    \mathbf{x}_j=\mathbf{W}_j\mathbf{s}_j,
\end{equation}
where $\mathbf{s}_j\sim \mathcal{CN}(\mathbf{0}_{N\times 1},\bar{P}_j\mathbf{I}_N)$ is the information-bearing signal vector. $\mathbf{W}_j\in\mathbb{C}^{M \times N}$ denotes the normalized precoding matrix with
$\textmd{Tr}\{\mathbf{W}_j\mathbf{W}_j^{\dag}\}=1$.
The total transmit power of each BS is assumed to be fixed at $P_t$, and the power is equally divided over users, i.e., $\bar{P}_j=\frac{P_t}{K}$, for all $j\in \mathcal{K}_i$, $i=0,\cdots,6$.

The second and the third terms on the right-hand side of (\ref{general system model}), i.e., $\mathbf{u}_k^{intra}=\sum_{j\in\mathcal{K}_0,j\neq k}\mathbf{G}_{k,\mathcal{B}_0}\mathbf{x}_j$ and $\mathbf{u}_k^{inter}=\sum_{i=1}^6\sum_{j\in\mathcal{K}_i}\mathbf{G}_{k,\mathcal{B}_i}\mathbf{x}_j$, denote the intra-cell interference and inter-cell interference received at user $k$, respectively. With a large number of BS antennas $M\gg 1$, $\mathbf{u}_k^{intra}$ and $\mathbf{u}_k^{inter}$ can be modeled as complex Gaussian random vectors with zero mean and covariance matrices $\mathbf{Q}_{k}^{intra}$ and $\mathbf{Q}_{k}^{inter}$, respectively.
Note that the transmitted signal $\mathbf{x}_j$ for user $j\in \mathcal{K}_i$ is independent of the channel gain matrix $\mathbf{G}_{k,\mathcal{B}_i}$ from BS $i$ to user $k\in\mathcal{K}_0$, $i=1,\dots,6$. For a large number of users $K\gg 1$, Appendix \ref{app inter} shows that the covariance matrix $\mathbf{Q}_k^{inter}$ of inter-cell interference of user $k$ can be obtained as
\begin{equation}\label{intercell}
    \mathbf{Q}_{k}^{inter}=\frac{1}{M}\sum_{i=1}^6 \sum_{m\in\mathcal{B}_i} |\gamma_{k,m}|^2 P_t \mathbf{I}_N.
\end{equation}

In this paper, we normalize the total system bandwidth to unity and focus on the spectral efficiency. According to (\ref{general system model}) and (\ref{TransVec}), the maximum achievable ergodic rate of user $k$ can be written as $\tilde{R}_k=NR_k$, where the per-antenna rate $R_k$ is given by
\begin{equation}\label{CapacityInGeneral}
R_k{=}\frac{1}{N}\mathbb{E}_{{\mathbf{H}}_{k,\mathcal{B}_0}}\left[\log_2\det\left(\mathbf{I}_N{+}\frac{\bar{P}_{k}\|\bm{\gamma}_{k,\mathcal{B}_0}\|^2
{{\mathbf{\tilde{G}}}_{k,{\mathcal{B}_0}}}{{\mathbf{W}}_{k}}{{\mathbf{W}}_{k}^\dag}
{{\mathbf{\tilde{G}}}_{k,{\mathcal{B}_0}}^\dag} }{N_0\mathbf{I}_N+\mathbf{Q}_k^{intra}+\mathbf{Q}_{k}^{inter}}\right)\right].
\end{equation}
${{\mathbf{\tilde{G}}}_{k,{\mathcal{B}_0}}}$ is the normalized channel gain matrix, which is defined as
\begin{equation}\label{tildeG def}
    {{\mathbf{\tilde{G}}}_{k,{\mathcal{B}_0}}}=\mathbf{B}_{k,\mathcal{B}_0} \circ \mathbf{H}_{k,\mathcal{B}_0},
\end{equation}
where $\mathbf{B}_{k,\mathcal{B}_0}\in \mathbb{C}^{N\times M}$ is the normalized large-scale fading matrix, which is composed of $N$ identical row vectors $\bm{\beta}_{k,\mathcal{B}_0}$ with entries
\begin{equation}\label{beta def}
    \beta_{k,m}=\frac{\gamma_{k,m}}{\|\bm{\gamma}_{k,\mathcal{B}_0}\|},
\end{equation}
for $m\in\mathcal{B}_0$. It is clear from (\ref{beta def}) that for any user $k\in\mathcal{K}_0$, $\|\bm{\beta}_{k,\mathcal{B}_0}\|=1$.

\begin{figure*}
\centerline{\subfloat[]{\includegraphics[width=3in]{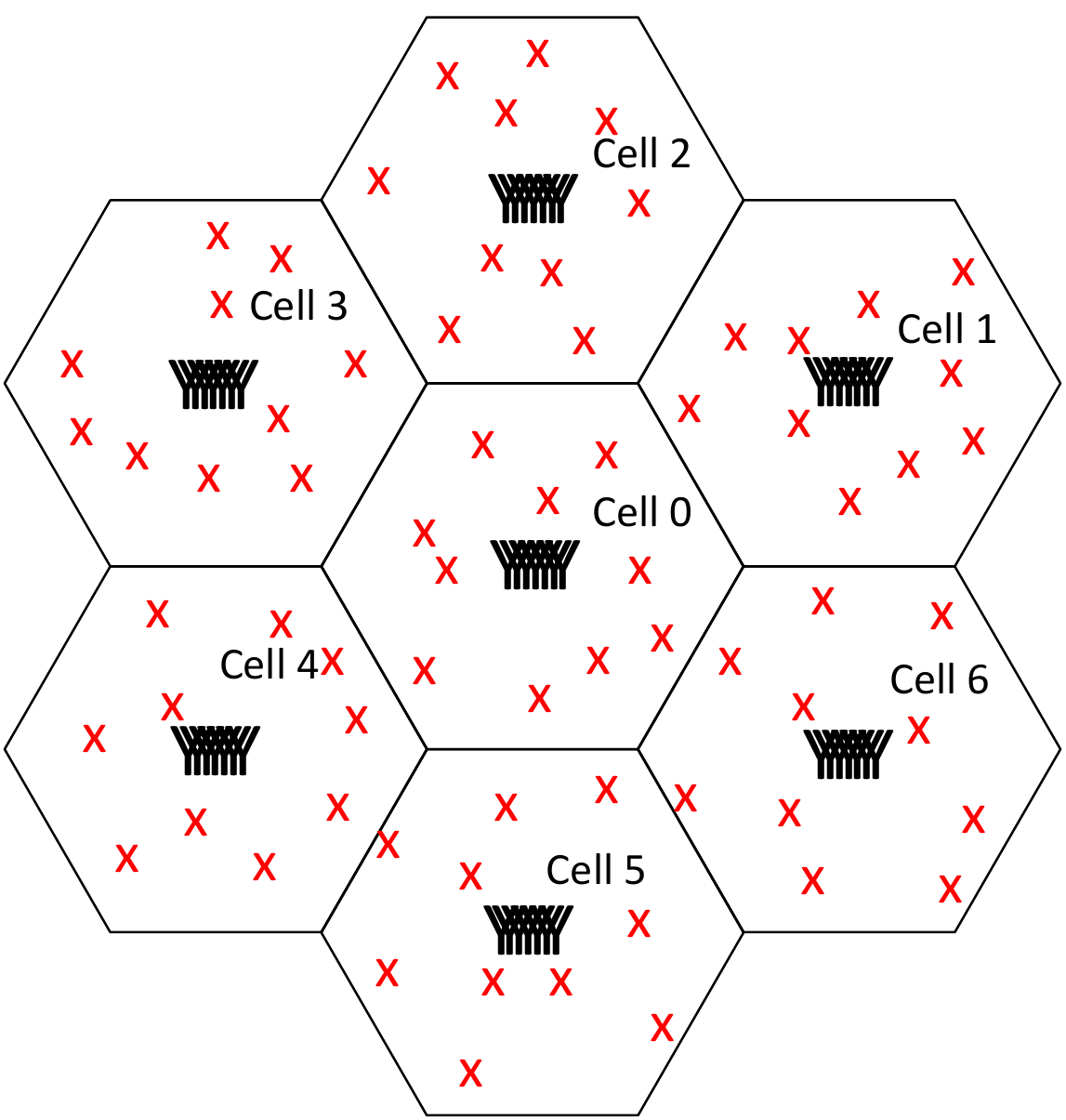}
\label{FIG_CA}}\hfil
\subfloat[]{\includegraphics[width=3.1in]{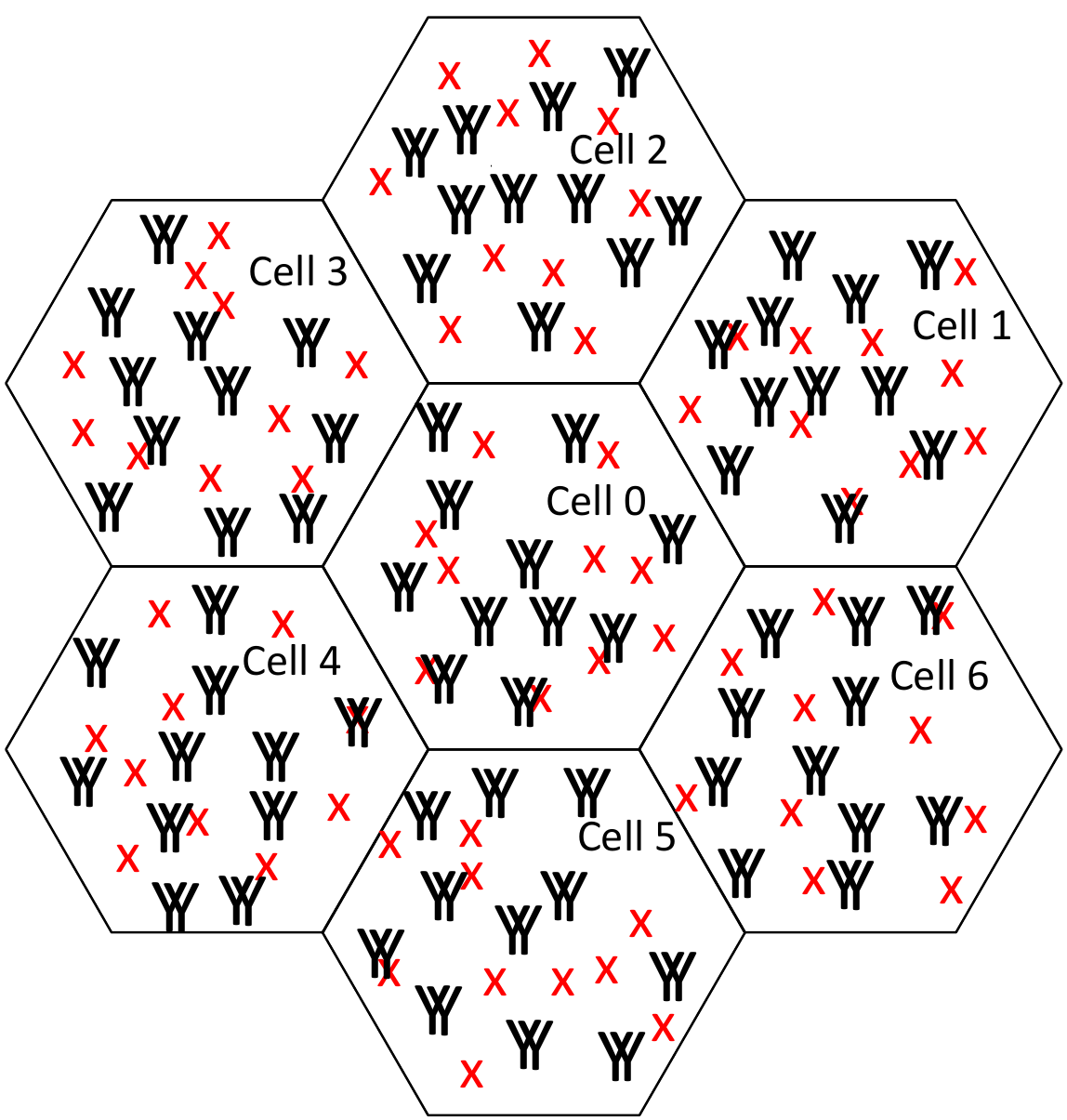}
\label{FIG_DA}}} \caption{A 1-tier hexagonal cellular network with $K$ uniformly distributed users in each cell and $M$ BS antennas with two antenna layouts: (a) with the CA layout, BS antennas are co-located at the center of each cell, and (b) with the DA layout, BS antennas are grouped as a set of antenna clusters that are uniformly distributed in  the inscribed circle of each cell. "Y" represents a BS antenna and "x" represents a user.}
\label{FIG_antenna}
\end{figure*}

(\ref{CapacityInGeneral}) indicates that the per-antenna rate $R_k$ is closely dependent on the large-scale fading vector $\bm{\gamma}_{k,\mathcal{B}_0}$. In this paper, we ignore the shadowing effect and model the large-scale fading coefficient of user $k$ to BS antenna $m$ as
\begin{equation} \label{gamma km}
\gamma_{k,m} =\left\|\mathbf{r}_{m}^{B} -\mathbf{r}_{k}^{U} \right\|^{-\alpha/2} ,
\end{equation}
where $\alpha>2$ is the path-loss factor. $\mathbf{r}_{k}^{U}$ and $\mathbf{r}_{m}^{B}$ denote the position of user $k$ and the position of BS antenna $m$, respectively. It is clear from (\ref{gamma km}) that the large-scale fading coefficients vary with the positions of users and BS antennas. In this paper, we assume that $K$ users are uniformly distributed in the inscribed circle of each hexagonal cell, and consider two BS antenna layouts as shown in Fig. \ref{FIG_antenna}: (a) the co-located antenna (CA) layout where $M$ BS antennas are placed at the center of each cell, and (b) the distributed antenna (DA) layout where $M$ BS antennas in each cell are grouped into $\tfrac{M}{N}$ clusters with $N$ BS antennas in each cluster. Denote the set of BS antennas of the $l$-th  cluster in Cell $i$ as $\mathcal{L}_l^i$. We have $|\mathcal{L}_l^i|=N$, $l=1,\cdots,L$, $i=0,\cdots,6$. The clusters are supposed to be uniformly distributed in the inscribed circle of each hexagonal cell.

It is clear from (\ref{gamma km}) that the large-scale fading coefficients of user $k$ depend on its distances to BS antennas. With the CA layout, the positions of BS antennas are given by
\begin{equation}\label{BS CA}
{\mathbf{r}}_m^B = \left\{ {\begin{array}{*{20}{c}}
   {(0,0)} & {m \in {\mathcal{B}_0}}  \\
   {(2,i \cdot \frac{\pi }{3} - \frac{\pi }{6})} & {m \in {\mathcal{B}_i},i = 1, \cdots ,6}.  \\
\end{array} } \right.
\end{equation}
For user $k\in\mathcal{K}_0$ at $(\rho_k,\theta_k)$, its large-scale fading coefficient $\gamma_{k,m}^C$ can be obtained by combining (\ref{gamma km}) and (\ref{BS CA}) as
\begin{equation}\label{gamma CA}
\gamma_{k,m}^C = \left\{ {\begin{array}{*{20}{c}}
   {\rho_k^{-\alpha/2}} & {m \in {\mathcal{B}_0}}  \\
   {\left(\rho_k^2+4-4\rho_k\cos\left(\theta_k-\left(i\cdot \frac{\pi}{3}-\frac{\pi}{6}\right)\right)\right)^{-\alpha/4}} & {m \in {\mathcal{B}_i},i = 1, \cdots ,6.}  \\
\end{array} } \right.
\end{equation}

With the DA layout, the BS antennas are grouped into clusters in each cell. The large-scale fading coefficient of user $k\in\mathcal{K}_0$ to BS antenna $m$ can be then written as
\begin{equation}\label{gamma DA}
    \gamma_{k,m}^D=d_{k,l,i}^{-\alpha/2},
\end{equation}
for $m\in\mathcal{L}_l^i$, where $d_{k,l,i}$ denotes the distance from user $k$ to BS antenna cluster $l$ in Cell $i$, $l=1,\cdots,L$, $i=0,\cdots,6$. With BS antenna clusters uniformly distributed in the inscribed circle of each cell, \cite{Dai2011} shows that the access distance $d_{k,l,0}$ given the position of user $k\in\mathcal{K}_0$ at $(\rho_k,\theta_k)$ has the following conditional cumulative distribution function (cdf) and probability density function (pdf) as
\begin{equation}\label{cdf d0}
{F_{d_{k,l,0}|{\rho_k}}}(x|y) {=} \left\{ {\begin{array}{*{20}{c}}
   {x^2} & \hspace{-0.3cm}{0 {\leq} x {\leq} 1 {-} y}  \\
   {   x^2(1 - \frac{1}{\pi }\arccos \frac{{1- {x^2} - {y^2}}}{{2xy}}) + \frac{1}{\pi }\arccos \frac{{1 - {x^2} + {y^2}}}{{2y}} - \frac{2}{{\pi }}{S_\Delta }
    } & {1 {-} y {<} x {\leq} 1 {+} y}  \\
\end{array} } \right.
\end{equation}
with
\begin{equation}\label{sdelta}
S_{\Delta}{=}\sqrt {\tfrac{{1 + x + y}}{2}{\left( {\tfrac{{1 + x + y}}{2}{-}1} \right)}{\left( {\tfrac{{1 + x + y}}{2}{-}x} \right)}{\left( {\tfrac{{1 + x + y}}{2}{-}y} \right)}},
\end{equation}
and
\begin{equation}\label{pdf d0}
{f_{{d_{k,l,0}}|{\rho_k}}}(x|y) = \left\{ {\begin{array}{*{20}{c}}
   {2x} & {0 {\leq} x {\leq} 1 {- }y}  \\
   {\frac{{2x}}{\pi }\arccos \frac{{{x^2} + {y^2} - {1}}}{{2xy}}} & {1 {- }y {<} x {\leq} 1 {+} y,}  \\
\end{array} } \right.
\end{equation}
respectively. For the distance $d_{k,l,i}$ from user $k\in\mathcal{K}_0$ to BS antenna cluster $l$ in Cell $i$, $i=1\cdots,6$, Appendix \ref{app pdf d} shows that its conditional pdf given the position of user $k$ at $(\rho_k,\theta_k)$ is given by
\begin{equation}\label{pdf d}
    {f_{{d_{k,l,i}}|{\rho _k},{\theta _k}}}(x|y,z)=\frac{{2x}}{\pi }\arccos \frac{{x^2+{{y^2} + 3 - 4y\cos \left( {z - \left( {i\cdot\tfrac{\pi }{3} - \tfrac{\pi }{6}} \right)} \right)} }}{{2x\sqrt {{y^2} + 4 - 4y\cos \left( {z - \left( {i\cdot\tfrac{\pi }{3} - \tfrac{\pi }{6}} \right)} \right)} }},
\end{equation}
if
\begin{equation}\label{pdf d range}
\sqrt {{y^2} + 4 - 4y\cos \left( {z - \left( {i\cdot\frac{\pi }{3} - \frac{\pi }{6}} \right)} \right)}  - 1\leq x\leq \sqrt {{y^2} + 4 - 4y\cos \left( {z - \left( {i\cdot\frac{\pi }{3} - \frac{\pi }{6}} \right)} \right)}  + 1.
\end{equation}
Otherwise ${f_{{d_{k,l,i}}|{\rho _k},{\theta _k}}}(x|y,z)=0$, $i=1,\cdots,6$. In contrast to $d_{k,l,0}$ which only depends on user $k$'s radial coordinate $\rho_k$, $d_{k,l,i}$ is further determined by its angular coordinate $\theta_k$, $i=1,\cdots,6$.

It is clear from (\ref{CapacityInGeneral}) and (\ref{gamma CA}-\ref{pdf d range}) that the per-antenna rate $R_k$ is determined by the positions of user $k$ and BS antennas. To study the scaling behavior of the per-antenna rate, we further define the average per-antenna rate $\bar{R}$ as
\begin{equation}\label{Cav def}
    \bar{R}\triangleq \mathbb{E}_{\mathbf{r}_k^U,\{\mathbf{r}_m^B\}_{m\in\mathcal{B}_i,i=0,\cdots,6}}\left[R_k\right],
\end{equation}
where the per-antenna rate $R_k$ is averaged over all possible positions of user $k$ and BS antennas. Note that with the CA layout, the positions of BS antennas are given in (\ref{BS CA}). The average per-antenna rate with the CA layout $\bar{R}^C$ is then reduced to
\begin{equation}\label{Cav def ca}
    \bar{R}^C\triangleq \mathbb{E}_{\mathbf{r}_k^U}\left[R_k^C\right].
\end{equation}

In this paper, we focus on the effect of BS antenna layout on the average rate performance when the number of BS antennas is large. In the following sections, an asymptotic analysis will be presented by assuming that the number of BS antennas $M$ and the number of user antennas $N$ go to infinity with $M/N\to L\gg K$. Note that with the DA layout, because $M=NL$, the assumption is simplified to $N\to\infty$.

\section{Single-User Capacity}

For illustration, let us start from the single-user case, i.e., $K=1$. Specifically, assume that a single user is randomly located in Cell 0, and its position follows a uniform distribution in the inscribed circle of Cell 0.
According to \cite{Telatar1999}, the capacity can be achieved by the singular-value-decomposition (SVD) transmission, and the corresponding precoding matrix $\mathbf{W}_k^{SVD}$ is given by
\begin{equation}\label{W su}
    \mathbf{W}_{k}^{SVD}=\mathbf{V}_{k,\mathcal{B}_0}\mathbf{\Omega}_{k}.
\end{equation}
$\mathbf{V}_{k,\mathcal{B}_0}$ is a unitary matrix obtained from the SVD of the normalized channel gain matrix $\mathbf{\tilde{G}}_{k,\mathcal{B}_0}$:
\begin{equation}\label{SVD of SU}
\mathbf{\tilde{G}}_{k,\mathcal{B}_0}=\mathbf{U}_{k,\mathcal{B}_0}\mathbf{\Lambda}_{k,\mathcal{B}_0}\mathbf{V}_{k,\mathcal{B}_0}^\dag,
\end{equation}
where ${\mathbf{\Lambda}}_{k,\mathcal{B}_0}=\left[\textrm{diag}\left(\sqrt{\lambda_1},\sqrt{\lambda_2},\dots,\sqrt{\lambda_N}\right),\mathbf{0}_{N\times (M-N)}\right]$ is composed by eigenvalues $\{\lambda_n\}$ of  $\mathbf{\tilde{G}}_{k,\mathcal{B}_0}\mathbf{\tilde{G}}_{k,\mathcal{B}_0}^\dag$.
$\mathbf{\Omega}_k$ denotes the power distribution over $N$ parallel sub-channels, which is given by
\begin{equation}\label{Omega_k}
\mathbf{\Omega}_{k}=\left[\textrm{diag}\left(
\sqrt{\frac{P_k(\lambda_1)}{\bar{P}_k\|\bm\gamma_{k,\mathcal{B}_0}\|^2}},\sqrt{\frac{P_k(\lambda_2)}{\bar{P}_k\|\bm\gamma_{k,\mathcal{B}_0}\|^2}},
\dots,\sqrt{\frac{P_k(\lambda_N)}{\bar{P}_k\|\bm\gamma_{k,\mathcal{B}_0}\|^2}}\right),\mathbf{0}_{N\times (M-N)}\right]^T,
\end{equation}
with $\{P_k(\lambda_n)\}$ denoting the water-filling power allocation, i.e.,
\begin{equation}\label{Waterfilling}
    P_k(\lambda_n)=\left(\zeta-\frac{N_0}{\lambda_n}\right)^+,
\end{equation}
where $(x)^+=\max(x,0)$, and $\zeta$ is chosen to satisfy
\begin{equation}\label{waterfilling_constraint}
    \sum_{n=1}^N P_k(\lambda_n)=\bar{P}_k\|\bm\gamma_{k,\mathcal{B}_0}\|^2.
\end{equation}
By combining (\ref{W su}-\ref{waterfilling_constraint}) with (\ref{CapacityInGeneral}) and ignoring the intra-cell interference and inter-cell interference terms, the single-user per-antenna capacity $R_k^S$ can be obtained as
\begin{equation}\label{Ck su}
    R_k^S=\frac{1}{N}\mathbb{E}_{\mathbf{H}_{k,\mathcal{B}_0}}\left[\log_2\det\left(\mathbf{I}_N+\mu_k\mathbf{\Lambda}_{k,\mathcal{B}_0}\mathbf{\Omega}_{k}
    \mathbf{\Omega}_{k}^\dag\mathbf{\Lambda}_{k,\mathcal{B}_0}^\dag\right)\right],
\end{equation}
where $\mu_k$ denotes the average per-antenna received signal-to-noise ratio (SNR), which is given by
\begin{equation}\label{mu su}
    \mu_k=\frac{\bar{P}_k\|\bm\gamma_{k,\mathcal{B}_0}\|^2}{N_0}.
\end{equation}


\subsection{Asymptotic Average Capacity with the CA Layout}

With the CA layout, all the BS antennas are placed at the center of the cell. By combining (\ref{gamma CA}) and (\ref{mu su}), the average per-antenna received SNR can be obtained as
\begin{equation}\label{mu su ca}
    \mu^C_k=\frac{M\bar{P}_k\rho_k^{-\alpha}}{N_0}.
\end{equation}
Moreover, according to (\ref{tildeG def}-\ref{beta def}) and (\ref{gamma CA}), the normalized channel gain matrix with the CA layout is given by $\mathbf{\tilde{G}}_{k,\mathcal{B}_0}^C{=}{\sqrt{\tfrac{1}{M}}}\mathbf{H}_{k,\mathcal{B}_0}$.
As $M,N\to\infty$ with $M/N\to L\geq 1$, the empirical eigenvalue distribution of $\mathbf{\tilde{G}}_{k,\mathcal{B}_0}^C\left(\mathbf{\tilde{G}}_{k,\mathcal{B}_0}^{C}\right)^{\dag}$
${\sim } \mathcal{W}_N(M,\tfrac{1}{M}\mathbf{I}_N)$ converges almost surely to the following distribution\cite{Marcenko1967}:
\begin{equation}\label{dis_ca su}
f_{\lambda}(x)= \left\{ \begin{array}{c} {\frac{1}{2\pi x}\sqrt{(x_{+}-L x)(L x-x_{-})}} \\ {0} \end{array}
\right. \begin{array}{cc} {} & {\textmd{if }\frac{1}{L}x_{-} \le x\le \frac{1}{L}x_{+}} \\ {} & {\textmd{otherwise,}} \end{array}
\end{equation}
where $x_+=\left(\sqrt{L} +1\right)^{2} $ and $x_{-}=\left(\sqrt{L} -1\right)^{2}$. As $L$ grows, the eigenvalues of $\mathbf{\tilde{G}}_{k,\mathcal{B}_0}^C\left(\mathbf{\tilde{G}}_{k,\mathcal{B}_0}^C\right)^{\dag}$ become increasingly deterministic, and eventually converge to $\mathbb{E}[\lambda]=1$. As a result, we have $\mathbf{\Lambda}_{k,\mathcal{B}_0}^C\approx \left[\mathbf{I}_N,\mathbf{0}_{N\times (M-N)}\right]$
for large $L\gg 1$. As $M,N\to\infty$ and $M/N\to L \gg 1$, the asymptotic per-antenna capacity with the CA layout can be then obtained by combining (\ref{Ck su}) and (\ref{mu su ca}) as\footnote{Note that for small $L$, (\ref{Capacity SU CA approx}) serves as a close upper-bound for the asymptotic per-antenna capacity.}
\begin{equation}\label{Capacity SU CA approx}
    R_k^{S-C}\approx \log_2 \left(1+L\frac{\bar{P}_k}{N_0}\rho_k^{-\alpha}\right).
\end{equation}

As we can see from (\ref{Capacity SU CA approx}), the asymptotic per-antenna capacity with the CA layout $R_{k}^{S-C}$ varies with the radial coordinate of the user $\rho_k$. By combining (\ref{Cav def ca}) and (\ref{Capacity SU CA approx}), the asymptotic average per-antenna capacity with the CA layout can be obtained as
\begin{equation}\label{Cav ca}
    \bar{R}^{S-C}=\int_0^1\log_2 \left(1+L\frac{\bar{P}_k}{N_0}x^{-\alpha}\right)f_{\rho_k}(x) dx,
\end{equation}
where $f_{\rho_k}(x)=2x$ is the pdf of the radial coordinate $\rho_k$ of user $k$. For large $L\gg 1$, we have
\begin{equation}\label{ca order}
\bar{R}^{S-C}\approx \log_2 \left(\frac{\bar{P}_k}{N_0}\right)  +\frac{\alpha}{\ln4} + \log_2 L.
\end{equation}

\subsection{Asymptotic Average Capacity with the DA Layout}

With the DA layout, both the average per-antenna received SNR $\mu_k^D$ and the eigenvalue distribution of $\mathbf{\tilde{G}}_{k,\mathcal{B}_0}^D\left(\mathbf{\tilde{G}}_{k,\mathcal{B}_0}^D\right)^\dag$ depend on the positions of BS antennas. By assuming no CSIT, i.e., equal power allocation among BS antennas, the asymptotic capacity for given user's and BS antennas' positions was derived as an implicit function of the large-scale fading coefficients from the user to $L$ BS antenna clusters \cite{Heliot2011,Aktas2006,Feng2009b,Zhang2013b}. Yet how the average capacity scales with $L$ remains largely unknown. In this section, we resort to an asymptotic lower-bound to study the scaling behavior of the single-user average capacity with the DA layout.

In particular, Appendix \ref{app su} shows that the per-antenna capacity with the DA layout $R_k^{S-D}$ is lower-bounded by
\begin{equation}\label{Capacity su da 1}
    R^{S-D}_{k,lb}{=}\frac{1}{N}\mathbb{E}_{\mathbf{H}_{k,0}^{(1)}}\left[\log_2\det\left(\mathbf{I}_N+\frac{1}{N}\frac{\bar{P}_k}{N_0} \left(d_{k,0}^{(1)}\right)^{-\alpha}\mathbf{H}_{k,0}^{(1)}\left(\mathbf{H}_{k,0}^{(1)}\right)^{\dag}\right)\right],
\end{equation}
where $d_{k,0}^{(1)}$ and $\mathbf{H}^{(1)}_{k,0}\in \mathbb{C}^{N\times N}$ denote the access distance from the user to its closest antenna cluster and the corresponding small-scale fading matrix, respectively.
As $N\to\infty$, the empirical eigenvalue distribution of $\frac{1}{N}\mathbf{H}^{(1)}_{k,0}\left(\mathbf{H}^{(1)}_{k,0}\right)^\dag$
$\sim  \mathcal{W}_N\left(N,\tfrac{1}{N}\mathbf{I}_N\right)$ converges almost surely to the following distribution\cite{Marcenko1967}:
\begin{equation}\label{dis_da su}
f_{{\lambda}}(x)=\left\{\begin{array}{c} {\frac{1}{2\pi x}\sqrt{4x-x^2}} \\ {0} \end{array}
\right. \begin{array}{cc} {} & {\textmd{if }0\le x\le 4} \\ {} & {\textmd{otherwise.}} \end{array}
\end{equation}
By combining (\ref{dis_da su}) and (\ref{Capacity su da 1}), the asymptotic lower-bound of the per-antenna capacity with the DA layout as $N\to\infty$ can be obtained as
\begin{equation}\label{Capacity su da 2}
    R^{S-D}_{k,lb}=\Phi\left(\frac{\bar{P}_k}{N_0} \left(d_{k,0}^{(1)}\right)^{-\alpha}\right),
\end{equation}
with
\begin{align}\label{Phi}
    \Phi(x)=2\log_2\left(\frac{1+\sqrt{1+4x}}{2}\right)-\frac{\log_2e}{4x}\left(\sqrt{1+4x}-1\right)^2 \mathop{\approx}\limits^{x\gg 1} \log_2x-\log_2e.
\end{align}
With $L\gg 1$, the minimum access distance $d_{k,0}^{(1)}\ll 1$. We then have
\begin{equation}\label{Capacity su da approx}
    R^{S-D}_{k,lb}\approx \log_2 \left(\frac{\bar{P}_k}{N_0}\left(d_{k,0}^{(1)}\right)^{-\alpha}\right) -\log_2 e.
\end{equation}

\begin{figure}
\begin{center}
\includegraphics[width=0.6\textwidth]{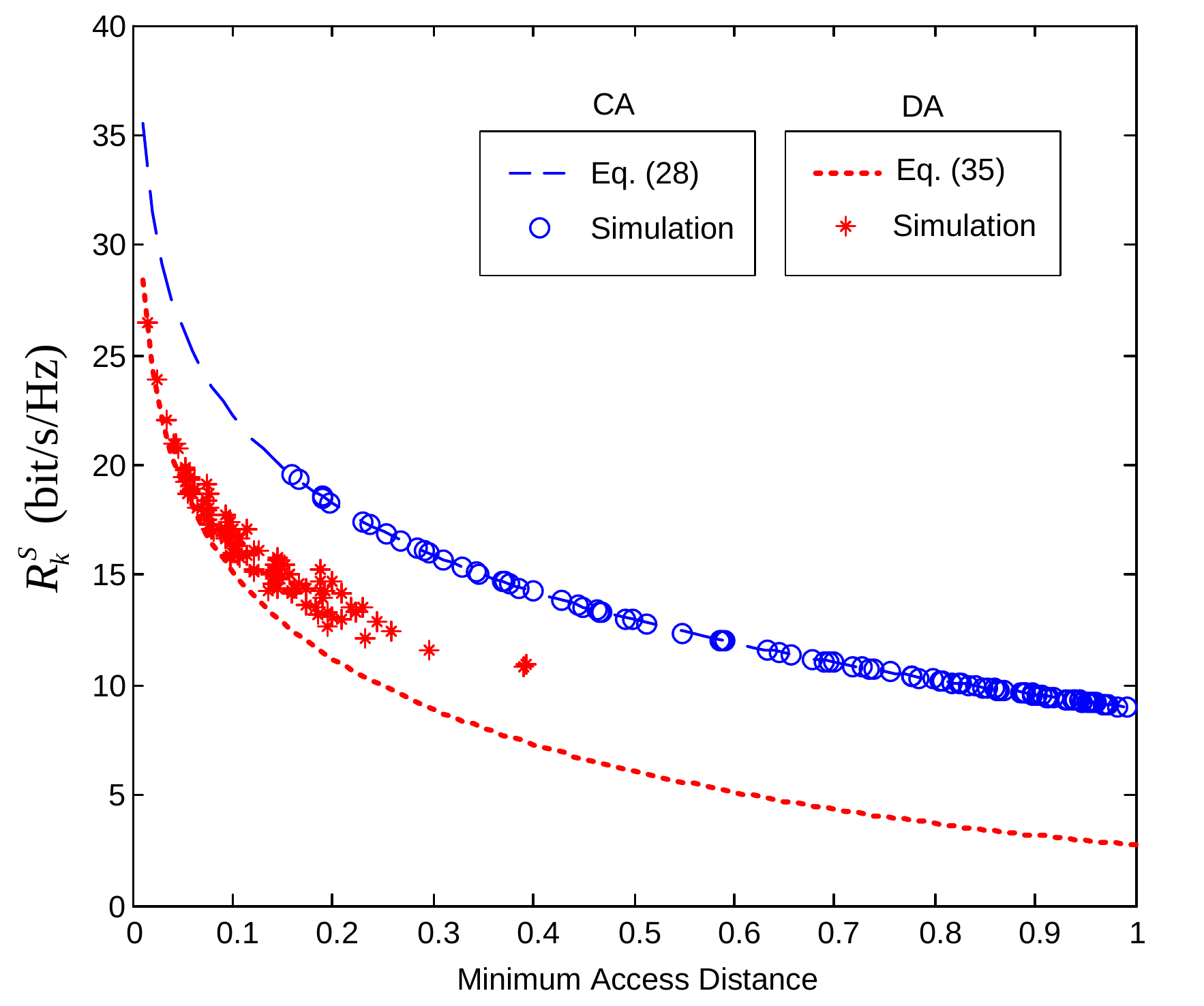}
\caption{Per-antenna capacity $R^S_k$ versus the minimum access distance of the user in the single-user case. $M=100$, $N=2$, $L=50$, $\alpha=4$, $\bar{P}_k/{N_0}=10$dB.}
\label{FIG_Cvsdist}
\end{center}
\end{figure}

We can see from (\ref{Capacity SU CA approx}) and (\ref{Capacity su da approx}) that both $R_k^{S-C}$ and $R^{S-D}_{k,lb}$ are crucially determined by the minimum access distance of the user.\footnote{With the CA layout, the minimum access distance is equal to the radial coordinate of user $k$, as all the BS antennas are co-located at the center of the cell.}
Fig. \ref{FIG_Cvsdist} plots the asymptotic per-antenna capacity  with the CA layout $R_k^{S-C}$ and the asymptotic lower-bound of the per-antenna capacity with the DA layout $R^{S-D}_{k,lb}$, where the minimum access distance in the x-axis is $\rho_k$ in the CA case and $d_{k,0}^{(1)}$ in the DA case, respectively. Simulation results of the per-antenna capacity with 100 realizations of the user's position are also presented. As we can see from Fig. \ref{FIG_Cvsdist}, the asymptotic capacity with the CA layout $R^{S-C}_{k}$ serves as a good approximation for the finite case even when the number of user antennas $N$ is small, i.e., $N=2$. With the DA layout, the asymptotic lower-bound $R^{S-D}_{k,lb}$ derived in (\ref{Capacity su da approx}) is found to be tight when the minimum access distance $d_{k,0}^{(1)}$ is small. Although for given minimum access distance, $R^{S-C}_{k}$ is always larger than $R^{S-D}_{k,lb}$, it can be observed from Fig. \ref{FIG_Cvsdist} that with the DA layout, the chance that the user has a small minimum access distance is much higher than that with the CA layout. We can then expect that a higher average per-antenna capacity could be obtained in the DA case thanks to the reduction of the minimum access distance.

By combining (\ref{Capacity su da approx}) and (\ref{Cav def}), the asymptotic lower-bound of the average per-antenna capacity with the DA layout $\bar{R}^{S-D}_{lb}$ can be further obtained as
\begin{align}\label{Capacity da lb approx}
    \bar{R}_{lb}^{S-D}= \int_0^1 \int_0^{1+y} \log_2 \left(x^{-\alpha}\right) f_{d_{k,0}^{(1)}|\rho_k} (x|y)f_{\rho_k}(y) dx dy +\log_2\left(\frac{\bar{P}_k}{N_0}\right)-\log_2 e,
\end{align}
where $f_{\rho_k}(y)=2y$ is the pdf of the radial coordinate $\rho_k$ of user $k$, and $f_{d_{k,0}^{(1)}|\rho_k} (x|y)$ is the conditional pdf of the minimum access distance $d_{k,0}^{(1)}$ of user $k$ given its position at ($\rho_k,\theta_k$), which is given by
\begin{equation}\label{pdf_dmin}
    f_{d_{k,0}^{(1)}|\rho_k}(x|y)=L(1-F_{d_{k,l,0}|\rho_k}(x|y))^{L-1}f_{d_{k,l,0}|\rho_k}(x|y),
\end{equation}
where $F_{d_{k,l,0}|\rho_k}(x|y)$ and $f_{d_{k,l,0}|\rho_k}(x|y)$ are given in (\ref{cdf d0}) and (\ref{pdf d0}), respectively.

\begin{figure}
\begin{center}
\includegraphics[width=0.6\textwidth]{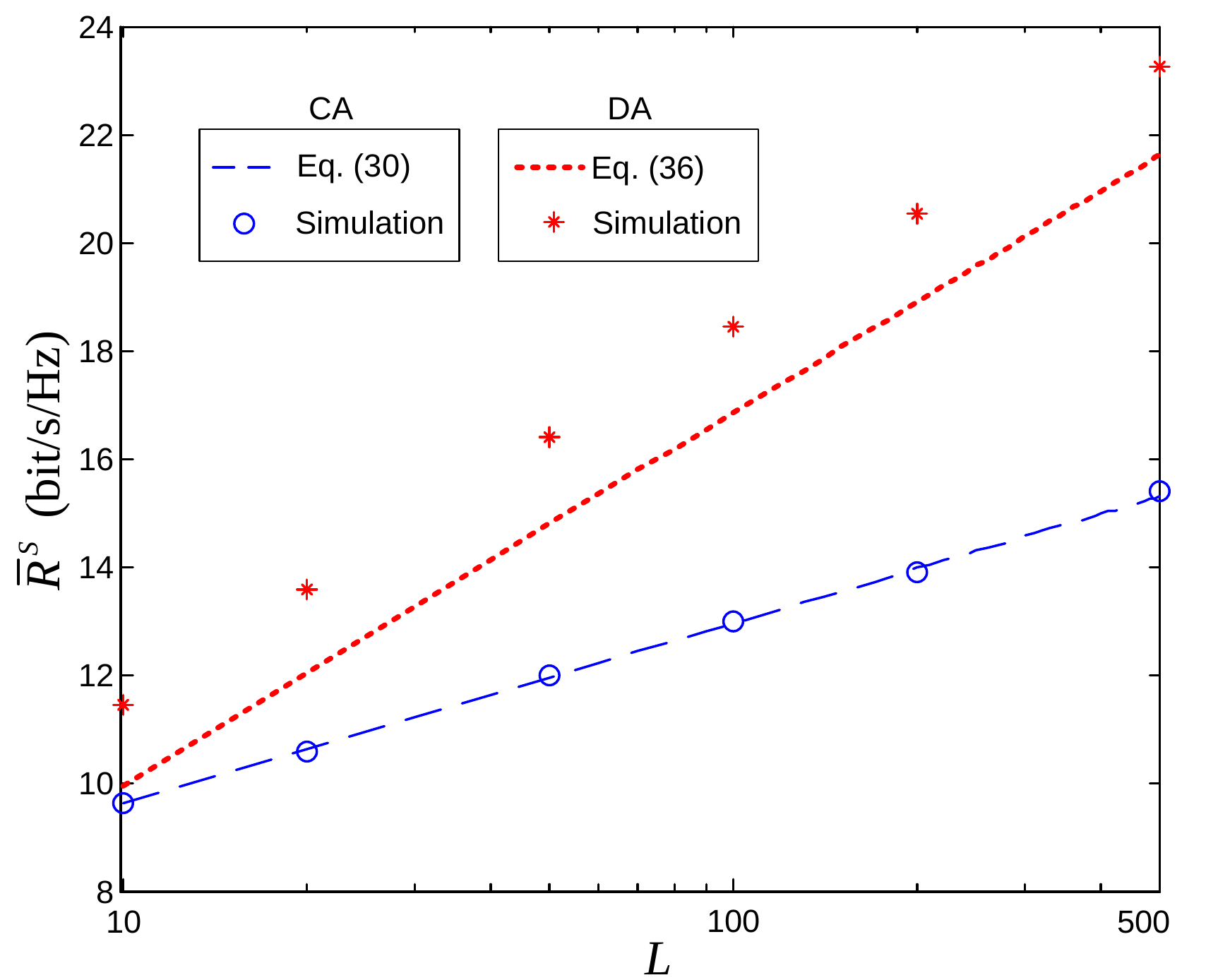}
\caption{Average per-antenna capacity $\bar{R}^S$ versus the ratio $L$ of the number of BS antennas $M$ to the number of user antennas $N$ in the single-user case. $N=2$, $\alpha=4$, $\bar{P}_k/N_0=10$dB. }
\label{FIG_su_L}
\end{center}
\end{figure}

The asymptotic average per-antenna capacity with the CA layout $\bar{R}^{S-C}$ and the asymptotic lower-bound of the average per-antenna capacity with the DA layout $\bar{R}^{S-D}_{lb}$ are plotted in Fig. \ref{FIG_su_L}. Intuitively, with $L$ uniformly distributed BS antenna clusters, the minimum access distance decreases in the order of $L^{-1/2}$ as $L$ increases. We can then see from (\ref{Capacity da lb approx}) that the asymptotic lower-bound $\bar{R}^{S-D}_{lb}$ increases in the order of $\tfrac{\alpha}{2}\log_2 L$, which is higher than $\bar{R}^{S-C}$ according to (\ref{ca order}) as the path-loss factor $\alpha>2$. It can be clearly observed from Fig. \ref{FIG_su_L} that $\bar{R}^{S-C}$ and $\bar{R}^{S-D}_{lb}$ logarithmically increase with $L$ in the orders of $\log_2 L$ and $\frac{\alpha}{2}\log_2 L$, respectively. $\bar{R}_{lb}^{S-D}$ is much higher than $\bar{R}^{S-C}$ when $L$ is large, indicating that substantial capacity gains can be achieved in the DA case when a large number of BS antennas are employed.

The analysis is verified by the simulation results of the average per-antenna capacity presented in Fig. \ref{FIG_su_L}. With the CA layout, the average per-antenna capacity is obtained by averaging over 100 realizations of the user's position. With the DA layout, it is further averaged over 100 realizations of the BS antenna topology. As we can see from Fig. \ref{FIG_su_L}, the average per-antenna capacities in both cases logarithmically increase with the ratio $L$ of the number of BS antennas $M$ to the number of user antennas $N$. Similar to its asymptotic lower-bound $\bar{R}_{lb}^{S-D}$, the average per-antenna capacity with the DA layout increases with $L$ in the order of $\frac{\alpha}{2}\log_2 L$, which is much higher than that with the CA layout when $L$ is large.

\section{Multi-User Rate with BD}
In Section III, we have shown that in the single-user case, the average per-antenna capacities with the CA and DA layouts both logarithmically increase with the ratio $L$ of the number of BS antennas $M$ to the number of user antennas $N$, but a higher scaling order is achieved in the DA case thanks to the reduction of the minimum access distance. With multiple users in each cell, users may suffer from interference from both intra-cell and inter-cell, which largely depends on the precoding strategy. In this section, we focus on a popular orthogonal linear precoding scheme, block diagonalization (BD) \cite{Spencer2004}, and study the effect of BS antenna layout on the scaling behavior of the average per-antenna rate with $K\gg 1$ users in each cell.

With BD, an intra-cell-interference-free block channel is obtained by projecting the desired signal to the null space of the channel gain matrices of the intra-cell users, and then decomposed to several parallel sub-channels. It requires that the number of BS antennas $M$ is no smaller than the total number of user antennas $KN$. In particular, for user $k\in\mathcal{K}_0$, define $\mathbf{X}_{k,\mathcal{B}_0}$ as
\begin{equation}\label{X}
    \mathbf{X}_{k,\mathcal{B}_0}=\left[{\mathbf{\tilde{G}}}_{1,\mathcal{B}_0}^T,\cdots,{\mathbf{\tilde{G}}}_{k-1,\mathcal{B}_0}^T,{\mathbf{\tilde{G}}}_{k+1,\mathcal{B}_0}^T,\cdots,{\mathbf{\tilde{G}}}_{K,\mathcal{B}_0}^T\right]^T,
\end{equation}
and denote its SVD as
\begin{equation}\label{X_SVD}
    {\mathbf{X}}_{k,\mathcal{B}_0}={\mathbf{\hat{U}}}_{k,\mathcal{B}_0}{\mathbf{\hat{\Lambda}}}_{k,\mathcal{B}_0}\left[{\mathbf{\hat{V}}}_{k,\mathcal{B}_0}^{(1)},{\mathbf{\hat{V}}}_{k,\mathcal{B}_0}^{(0)}\right]^\dag,
\end{equation}
where ${\mathbf{\hat{V}}}_{k,\mathcal{B}_0}^{(1)}$ holds the first $(K-1)N$ right singular vectors and ${\mathbf{\hat{V}}}_{k,\mathcal{B}_0}^{(0)}$ holds the rest. ${\mathbf{\hat{V}}}_{k,\mathcal{B}_0}^{(0)}$ corresponds to zero singular values and forms an orthogonal basis for the null space of $\mathbf{X}_{k,\mathcal{B}_0}$. Let $\mathbf{\tilde{X}}_{k,\mathcal{B}_0}=\mathbf{\tilde{G}}_{k,\mathcal{B}_0}{\mathbf{\hat{V}}}_{k,\mathcal{B}_0}^{(0)}$, and denote its SVD as
\begin{equation}\label{SVD of BD channel}
\mathbf{\tilde{X}}_{k,\mathcal{B}_0}={{\mathbf{\tilde{U}}}_{k,\mathcal{B}_0}}{{\mathbf{\tilde{\Lambda}}}_{k,\mathcal{B}_0}}{{\mathbf{\tilde{V}}}_{k,\mathcal{B}_0}^\dag},
\end{equation}
where ${\mathbf{\tilde{\Lambda}}}_{k,\mathcal{B}_0}{=}\left[\textrm{diag}\left(\sqrt{\tilde{\lambda}_1},\sqrt{\tilde{\lambda}_2},\dots,
\sqrt{\tilde{\lambda}_N}\right),\mathbf{0}_{N\times (M-KN)}\right]$ is composed by eigenvalues $\{\tilde{\lambda}_n\}$ of $\mathbf{\tilde{X}}_{k,\mathcal{B}_0}\mathbf{\tilde{X}}_{k,\mathcal{B}_0}^\dag$. The precoding matrix of user $k$ with BD can be written as \cite{Spencer2004}
\begin{equation}\label{W bd}
    \mathbf{W}_{k}^{BD}=\mathbf{\hat{V}}_{k,\mathcal{B}_0}^{(0)}\mathbf{\tilde{V}}_{k,\mathcal{B}_0}\mathbf{\tilde{\Omega}}_k,
\end{equation}
where $\mathbf{\tilde{\Omega}}_k$ denotes the power distribution of the $N$ parallel sub-channels, which is given by
\begin{equation}\label{Omega_k_BD}
\mathbf{\tilde{\Omega}}_{k}=\left[\textrm{diag}\left(\sqrt{\frac{P_k(\tilde{\lambda}_1)}{\bar{P}_k\|\bm{\gamma}_{k,\mathcal{B}_0}\|^2}},
\sqrt{\frac{P_k(\tilde{\lambda}_2)}{\bar{P}_k\|\bm{\gamma}_{k,\mathcal{B}_0}\|^2}},\cdots,\sqrt{\frac{P_k(\tilde{\lambda}_N)}{\bar{P}_k\|\bm{\gamma}_{k,\mathcal{B}_0}\|^2}}\right),\mathbf{0}_{N\times (M-KN)}\right]^T,
\end{equation}
with $\{P_k(\tilde{\lambda}_n)\}$ denoting the water-filling power allocation, i.e.,
\begin{equation}\label{Waterfilling_BD}
    P_k(\tilde{\lambda}_n)=\left(\tilde{\zeta}-\frac{N_0}{\tilde{\lambda}_n}\right)^+,
\end{equation}
where $\tilde{\zeta}$ is chosen to satisfy
\begin{equation}\label{waterfilling_constraint_BD}
    \sum_{n=1}^N P_k(\tilde{\lambda}_n)=\bar{P}_k\|\bm\gamma_{k,\mathcal{B}_0}\|^2.
\end{equation}

With BD, the intra-cell interference $\mathbf{u}_k^{intra}=\mathbf{0}$ as $\mathbf{\tilde{G}}_{k,\mathcal{B}_0}\mathbf{W}_j^{BD}=\mathbf{0}$ for all $j\in \mathcal{K}_0$ and $j\ne k$. By combining (\ref{CapacityInGeneral}) and (\ref{SVD of BD channel}-\ref{waterfilling_constraint_BD}), the per-antenna rate with BD of user $k\in\mathcal{K}_0$ can be obtained as
\begin{equation}\label{Rk bd}
R_k^M=\frac{1}{N}\mathbb{E}_{\mathbf{H}_{k,\mathcal{B}_0}}\left[\log_2\det\left(\mathbf{I}_N+\tilde{\mu}_k{\mathbf{\tilde{\Lambda}}}_{k,\mathcal{B}_0}
\mathbf{\tilde{\Omega}}_{k}\mathbf{\tilde{\Omega}}_{k}^\dag{\mathbf{\tilde{\Lambda}}}_{k,\mathcal{B}_0}^\dag\right)\right],
\end{equation}
where $\tilde{\mu}_k$ denotes the average received signal-to-interference-plus-noise ratio (SINR),
\begin{equation}\label{sinr}
    \tilde{\mu}_k=\frac{\frac{1}{K}\|\bm{\gamma}_{k,\mathcal{B}_0}\|^2}{\frac{N_0}{P_t}+P_k^{int}}.
\end{equation}
$P_k^{int}$ is the normalized inter-cell interference power, which can be obtained from (\ref{intercell}) as
\begin{equation}\label{Pk int def}
    P_k^{int}=\frac{1}{M}\sum_{i=1}^6\sum_{m\in\mathcal{B}_i}|\gamma_{k,m}|^2.
\end{equation}
(\ref{Rk bd}-\ref{sinr}) indicates that the per-antenna rate with BD closely depends on the normalized inter-cell interference power $P_k^{int}$, which, as shown in (\ref{Pk int def}), is determined by the large-scale fading coefficients between user $k$ and BS antennas in Cell $i$, $i=1,\cdots,6$. In the next section, we will examine how the normalized inter-cell interference power varies with different BS antenna layouts.

\subsection{Normalized Inter-cell Interference Power}
\subsubsection{CA}
For user $k\in\mathcal{K}_0$ at $(\rho_k,\theta_k)$, the normalized inter-cell interference power with the CA layout can be obtained by combining (\ref{Pk int def}) and (\ref{gamma CA}) as
\begin{equation}\label{inter ca}
     P_k^{int,C}=\sum_{i=1}^6\left(\rho_k^2+4-4\rho_k\cos\left(\theta_k-\left(i\cdot \frac{\pi}{3}-\frac{\pi}{6}\right)\right)\right)^{-\alpha/2}.
\end{equation}
(\ref{inter ca}) indicates that the normalized inter-cell interference power with the CA layout $P_k^{int,C}$ is solely determined by the position of user $k$. Due to the symmetric nature of the positions of BS antennas shown in (\ref{BS CA}), $P_k^{int,C}$ is a periodic function of period $\pi/3$ for any $\rho_k\in[0,1]$. It is maximized when $\theta_k=i\cdot\tfrac{\pi}{3}-\tfrac{\pi}{6}$, and minimized when $\theta_k=i\cdot\tfrac{\pi}{3}$, $i=1,\cdots,6$. For given $\theta_k$, $P_k^{int,C}$ is a monotonic increasing function with respect to $\rho_k\in[0,1]$. It can be easily shown that with the path-loss factor $\alpha=4$, $P_k^{int,C}$ is minimized at $(0,0)$ with $P_k^{int,C}|{(0,0)}=0.375$, and maximized at $(1,i\cdot\tfrac{\pi}{3}-\tfrac{\pi}{6})$ with $P_k^{int,C}|{(1,i\cdot\tfrac{\pi}{3}-\tfrac{\pi}{6})}\approx 1.275$, $i=1,\cdots,6$.

\subsubsection{DA}
With the DA layout, as $N$ BS antennas are co-located at each antenna cluster, the normalized inter-cell interference power $P_k^{int,D}$ can be obtained by combining (\ref{Pk int def}) and (\ref{gamma DA}) as
\begin{equation}\label{inter da}
    P_k^{int,D}=\frac{1}{L}\sum_{i=1}^6 \sum_{l=1}^L d_{k,l,i}^{-\alpha},
\end{equation}
with the $n$-th moment
\begin{equation}\label{nth moment inter}
    \mathbb{E}\left[\left(P_k^{int,D}\right)^n\right]=\frac{1}{L^n}\sum_{\sum_{l=1}^L t_{l}=n}\frac{n!}{\prod_{l=1}^Lt_{l}!}                                                                                 \prod_{l=1}^L\mathbb{E}\left[\left(\sum_{i=1}^6 d_{k,l,i}^{-\alpha}\right)^{t_{l}}\right],
\end{equation}
where the sum is taken over all possible combinations of nonnegative integers $t_{l}$ given $\sum_{l=1}^L t_{l}=n$. It is clear from (\ref{nth moment inter}) that the $n$-th moment of the normalized inter-cell interference $\mathbb{E}\left[\left(P_k^{int,D}\right)^n\right]$ crucially depends on the distribution of the distance $d_{k,l,i}$ from user $k\in\mathcal{K}_0$ to BS antenna cluster $l$ in Cell $i$, $l=1,\cdots,L$, $i=1,\cdots,6$, which varies with user $k$'s position as shown in (\ref{pdf d}-\ref{pdf d range}).

If user $k$ is at the cell center $(0,0)$, for instance, the conditional pdf of $d_{k,l,i}$ can be obtained from (\ref{pdf d}-\ref{pdf d range}) as
\begin{equation}\label{pdf center}
{f_{{d_{k,l,i}}|{\rho _k},{\theta _k}}}(x|0,0) = \left\{ {\begin{array}{*{20}{c}}
   {\frac{2x}{\pi}\arccos \frac{{{x^2} + 3}}{{4 x}}} & {\textrm{if } 1 \leq x \leq 3}  \\
   0 & {\textrm{otherwise},}  \\
\end{array} } \right.
\end{equation}
$i=1,\cdots,6$. We can see from (\ref{pdf center}) that ${f_{{d_{k,l,i}}|{\rho _k},{\theta _k}}}(x|0,0)$ is independent of $i$, indicating an isotropic normalized inter-cell interference power. With $\alpha=4$, the mean normalized inter-cell interference power for a cell-center user can be obtained by combining (\ref{nth moment inter}) and (\ref{pdf center}) as $\mathbb{E}\left[P_k^{int,D}|(0,0)\right]=\frac{2}{3}$, which is slightly higher than the normalized inter-cell interference power in the CA layout, i.e., $P_k^{int,C}|(0,0)=0.375$.

On the other hand, for a cell-edge user located at $(1,\tfrac{\pi}{6})$, the conditional pdf of $d_{k,l,i}$ can be obtained from (\ref{pdf d}-\ref{pdf d range}) as
\begin{equation}\label{pdf edge}
{f_{{d_{k,l,i}}|{\rho _k},{\theta _k}}}\left(x|1,\tfrac{\pi }{6}\right) = \left\{ {\begin{array}{*{20}{c}}
   {\frac{{2x}}{\pi }\arccos \frac{{4 - 4\cos \frac{{(1 - i)\pi }}{3} + {x^2}}}{{2x\sqrt {5 - 4\cos \frac{{(1 - i)\pi }}{3}} }}} & {\textrm{if } \sqrt {5 - 4\cos \frac{{(1 - i)\pi }}{3}}  - 1 \leq x \leq \sqrt {5 - 4\cos \frac{{(1 - i)\pi }}{3}}  + 1}  \\
   0 & {\textrm{otherwise,}}  \\
\end{array} } \right.
\end{equation}
$i=1,\cdots,6$. In this case, ${f_{{d_{k,l,i}}|{\rho_k},{\theta_k}}}(x|1,\tfrac{\pi}{6})$ varies with $i$, indicating
that the BS antenna clusters in different cells have distinct contributions to the normalized inter-cell interference power $P_k^{int,D}$. Specifically, as user $k$ is close to the neighboring Cell 1, we have $d_{k,l,1}\ll d_{k,l,i}$, for $i=2,\cdots,6$. As a result, $\sum_{i=1}^6 d_{k,l,i}^{-\alpha}\approx d_{k,l,1}^{-\alpha}$, and (\ref{nth moment inter}) reduces to
\begin{equation}\label{nth moment edge}
    \mathbb{E}\left[\left(P_k^{int,D}\right)^n|(1,\tfrac{\pi}{6})\right]\approx\frac{1}{L^n}\sum_{\sum_{l=1}^L t_{l}=n}\frac{n!}{\prod_{l=1}^Lt_{l}!}                                                                                 \prod_{l=1}^L\mathbb{E}\left[d_{k,l,1}^{-t_{l}\alpha}|(1,\tfrac{\pi }{6})\right].
\end{equation}
With $t_l\geq 1$ and the path-loss factor $\alpha>2$, we have $t_l\alpha>2$. $\mathbb{E}\left[ {d_{k,l,1}^{ - t_{l}\alpha }|(1,\tfrac{\pi }{6})} \right]$ can be then obtained from (\ref{pdf edge}) as
\begin{align}\label{app nth 1}
  &\mathbb{E}\left[ {d_{k,l,1}^{ - t_{l}\alpha }|(1,\tfrac{\pi }{6})} \right] = \frac{2}{\pi }\int_0^2 {{x^{ - t_{l}\alpha  + 1}}} \arccos \frac{x}{2}dx \nonumber \\
  & = \left\{ {\begin{array}{*{20}{c}}
   {\frac{1}{{\pi (t_{l}\alpha  - 2)(t_{l}\alpha  - 3)}}
   \left( {\frac{{{2^{3 - t_{l}\alpha }}\sqrt \pi  \Gamma (\frac{{5 - t_{l}\alpha }}{2})}}{{\Gamma (\frac{{4 - t_{l}\alpha }}{2})}}
    + \mathop {\lim }\limits_{x \to 0^+} \frac{{ 2(t_{l}\alpha  - 3)\arccos \frac{x}{2} - x{F_1}\left( {\frac{1}{2},\frac{{3 - t_{l}\alpha }}{2};\frac{{5 - t_{l}\alpha }}{2};\frac{{{x^2}}}{4}} \right)}}{{{x^{t_{l}\alpha  - 2}}}}} \right)}
     & { t_{l}\alpha  \ne 3}  \\
   {\frac{2}{\pi }\mathop {\lim }\limits_{x \to 0^+} \frac{1}{x}{\arccos \frac{x}{2}} - \frac{1}{2}\ln x } & { t_{l}\alpha  = 3}  \\
\end{array} } \right. \nonumber \\
&=\infty,
\end{align}
where $_2F_1(a,b;c;z)$ denotes the hypergeometric function. We can conclude from (\ref{nth moment edge}-\ref{app nth 1}) that for the cell-edge user at $(1,\tfrac{\pi}{6})$, the $n$-th moment of the normalized inter-cell interference power with the DA layout $\mathbb{E}\left[\left(P_k^{int,D}\right)^n|(1,\tfrac{\pi}{6})\right]=\infty$. Intuitively, the inter-cell interference power becomes extremely strong if the user is close to some BS antenna cluster in the neighboring cells. With BS antenna clusters uniformly distributed in each cell, there is a non-zero probability that some antenna cluster falls into the vicinity area of the user if it is located at the cell edge, thus leading to the divergence of inter-cell interference power.

\begin{figure}
\begin{center}
\includegraphics[width=0.6\textwidth]{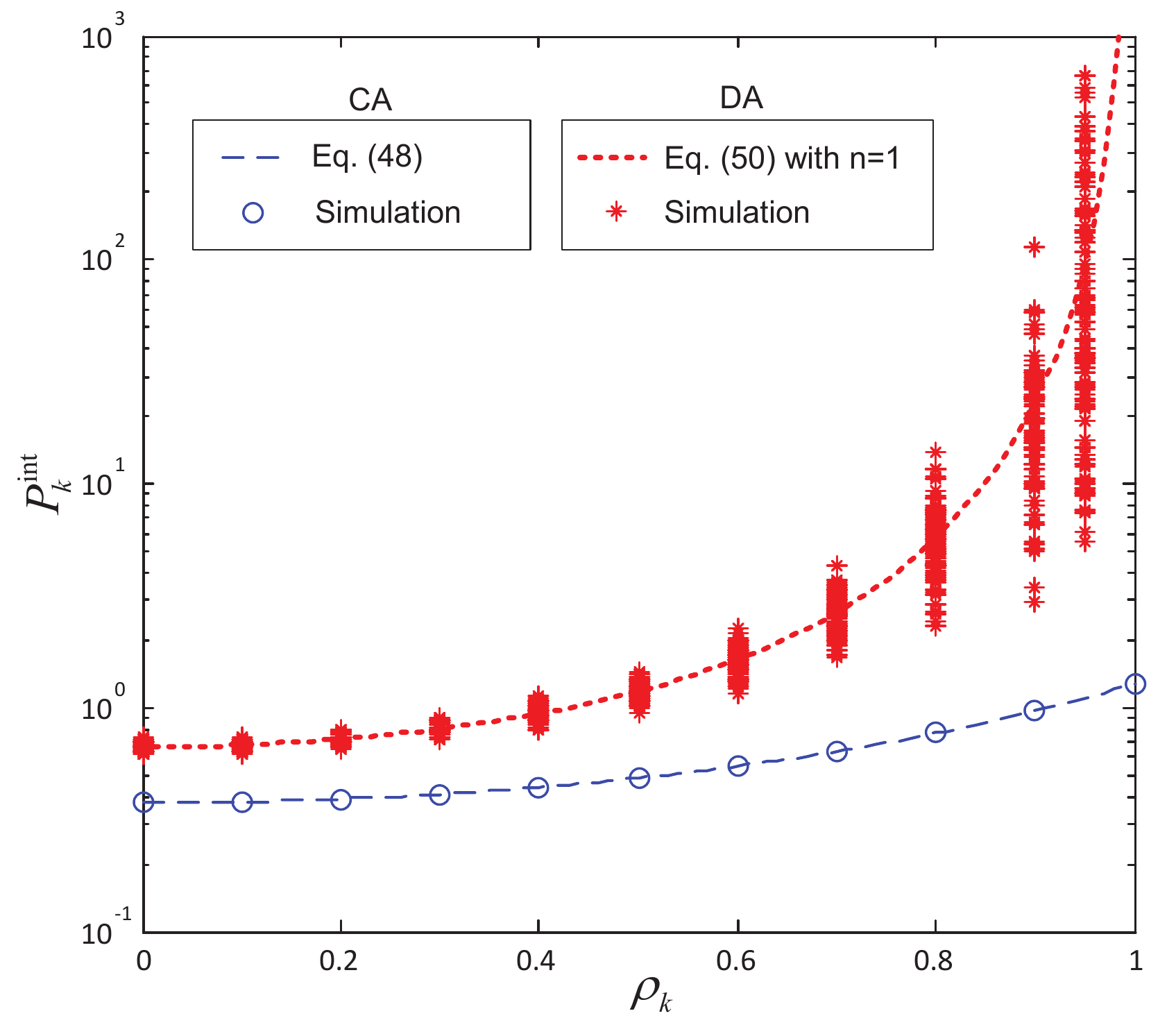}
\caption{Normalized inter-cell interference power $P_k^{int}$ of user $k$ versus its radial coordinate $\rho_k$. $\theta_k=\tfrac{\pi}{6}$. $K=50$, $L=200$, $N=2$, $M=400$ and $\alpha=4$. With the DA layout, simulation results are obtained based on 100 realizations of the BS antenna topology.}
\label{FIG_intercell}
\end{center}
\end{figure}

Fig. \ref{FIG_intercell} illustrates the normalized inter-cell interference power with the CA layout $P_k^{int,C}$ and the mean normalized inter-cell interference power with the DA layout $\mathbb{E}[P_k^{int,D}]$ of user $k$ given its angular coordinate at $\theta_k=\tfrac{\pi}{6}$. As we can see from Fig. \ref{FIG_intercell}, both $P_k^{int,C}$ and $\mathbb{E}[P_k^{int,D}]$ grow monotonically with the radial coordinate $\rho_k$ of user $k$ because of the reduction of the distances from user $k$ to BS antenna clusters in the neighboring Cell 1. With the DA layout, the mean normalized inter-cell interference power $\mathbb{E}[P_k^{int,D}]$ becomes infinite at $\rho_k=1$.

The analysis is verified by the simulation results presented in Fig. \ref{FIG_intercell}. It can be observed from Fig. \ref{FIG_intercell} that with the DA layout, in addition to the mean, the variance of the normalized inter-cell interference also grows with the radial coordinate $\rho_k$ of user $k$. It indicates that as the user moves towards the cell edge, the rate performance becomes increasingly sensitive to its position, as we will demonstrate in Section \ref{SectionIV-C}.

In the following sections, we will focus on the asymptotic average per-antenna rate as the number of BS antennas $M$ and the number of user antennas $N$ go to infinity with $M/N\to L\gg K$.\footnote{Note that BD requires that the number of BS antennas $M$ is no smaller than the total number of user antennas $KN$, or equivalently, $L\ge K$. Here we assume $L\gg K$ as the rate performance of BD is close to the downlink capacity when $M\gg N$, or equivalently, $L\gg K$.}

\subsection{Asymptotic Average Rate with the CA Layout}

According to (\ref{Rk bd}), the per-antenna rate with BD is crucially determined by the distribution of the eigenvalues of $\mathbf{\tilde{X}}_{k,\mathcal{B}_0}^C\left(\mathbf{\tilde{X}}_{k,\mathcal{B}_0}^C\right)^\dag$.
As $M,N\to\infty$ with $M/N\to L\geq K$, the empirical eigenvalue distribution of $\mathbf{\tilde{X}}_{k,\mathcal{B}_0}^C\left(\mathbf{\tilde{X}}_{k,\mathcal{B}_0}^C\right)^\dag\sim\mathcal{W}_N\left(M-(K-1)N,\tfrac{1}{M}\mathbf{I}_N\right)$ converges almost surely to the following distribution\cite{Marcenko1967}:
\begin{equation}\label{dis_eigenX}
f_{\tilde{\lambda}}(x)=\left\{\begin{array}{c} {\frac{1}{2\pi x}\sqrt{(\tilde{x}_{+}-Lx)(Lx-\tilde{x}_{-})}} \\ {0} \end{array}
\right. \begin{array}{cc} {} & {\textmd{if }\frac{1}{L}\tilde{x}_{-} \le x\le \frac{1}{L}\tilde{x}_{+}} \\ {} & {\textmd{otherwise,}} \end{array}
\end{equation}
where $\tilde{x}_{+} =\left(\sqrt{L-K+1} +1\right)^{2} $ and $\tilde{x}_{-} =\left(\sqrt{L-K+1} -1\right)^{2} $.
As $L$ grows, the eigenvalues of $\mathbf{\tilde{X}}_{k,\mathcal{B}_0}^C\left(\mathbf{\tilde{X}}_{k,\mathcal{B}_0}^C\right)^{\dag}$ become increasingly deterministic, and eventually converge to $\mathbb{E}[\tilde{\lambda}]=\tfrac{L-K+1}{L}$ \cite{Liu2012}. As a result, we have
$\mathbf{\tilde{\Lambda}}_{k,\mathcal{B}_0}^C\approx \left[\sqrt{1-\tfrac{K-1}{L}}\mathbf{I}_N,\mathbf{0}_{N\times (M-KN)}\right]$ for $L\gg K$. As $M, N\to\infty$ and $M/N\to L\gg K$, the asymptotic per-antenna rate can be obtained by combining (\ref{Rk bd}-\ref{inter ca}) as
\begin{equation}\label{rk bd ca}
    R_k^{M-C} \approx \log_2\left(1+\frac{L-K+1}{K}\cdot \frac{\rho_k^{-\alpha}}{\tfrac{N_0}{P_t}+\sum_{i=1}^6\left(\rho_k^2+4-4\rho_k\cos\left(\theta_k-\left(i\cdot \frac{\pi}{3}-\frac{\pi}{6}\right)\right)\right)^{-\alpha/2}}\right).
\end{equation}
(\ref{rk bd ca}) shows that the asymptotic per-antenna rate with the CA layout $R_k^{M-C}$ varies with the user's position $(\rho_k,\theta_k)$. By combining (\ref{rk bd ca}) and (\ref{Cav def ca}), the asymptotic average per-antenna rate with the CA layout can be further obtained as
\begin{align}\label{Rav bd ca scaling}
    \bar{R}^{M-C}&=\mathbb{E}_{\rho_k,\theta_k}\left[\log_2\left(1+\frac{L-K+1}{K}\cdot \frac{\rho_k^{-\alpha}}{\tfrac{N_0}{P_t}+\sum_{i=1}^6\left(\rho_k^2+4-4\rho_k\cos\left(\theta_k-\left(i\cdot \frac{\pi}{3}-\frac{\pi}{6}\right)\right)\right)^{-\alpha/2}}\right)\right]\nonumber \\
    &\approx\log_2\left(\frac{L}{K}-1\right) +\Psi^C(\alpha),
\end{align}
for $L\gg K\gg 1$ and $P_t/N_0\gg 1$, where $\Psi^C(\alpha)$ is given by
\begin{equation}\label{Psi}
        \Psi^C(\alpha)=\frac{\alpha}{\ln 4}-\frac{1}{\pi}\int_0^{2\pi}\int_0^1 x\log_2\left(\sum_{i=1}^6\left(x^2+4-4x\cos\left(y-\left(i\cdot \frac{\pi}{3}-\frac{\pi}{6}\right)\right)\right)^{-\alpha/2}\right) dx dy.
\end{equation}
With the path-loss factor $\alpha=4$, for instance, we have $\Psi^C(4)\approx 3.54$.

\subsection{Asymptotic Average Rate with the DA Layout} \label{SectionIV-C}

With the DA layout, the distribution of the eigenvalues of $\mathbf{\tilde{X}}_{k,\mathcal{B}_0}^D\left(\mathbf{\tilde{X}}_{k,\mathcal{B}_0}^D\right)^\dag$ is determined by the positions of BS antennas, which is difficult to characterize with BS antennas grouped into uniformly distributed clusters. Similar to the single-user case, we resort to a lower-bound to study the scaling behavior of the average per-antenna rate with the DA layout.

Specifically, Appendix \ref{app bd} shows that with $L\gg K$, the per-antenna rate with the DA layout $R_{k}^{M-D}$ is lower-bounded by
\begin{equation}\label{Rklb}
    R_{k,lb}^{M-D}=\frac{1}{N}\mathbb{E}_{\mathbf{\tilde{H}}_{k,0}^{(1)}}\left[\log_2\det\left(\mathbf{I}_N
    +\frac{\frac{1}{K}\left(\tilde{d}_{k,0}^{(1)}\right)^{-\alpha}}
    {\frac{N_0}{P_t}+\frac{1}{L}\sum_{i=1}^6\sum_{l=1}^Ld_{k,l,i}^{-\alpha}}\cdot\frac{1}{N}\mathbf{\tilde{H}}_{k,0}^{(1)}\left(\mathbf{\tilde{H}}_{k,0}^{(1)}\right)^{\dag}\right)\right],
\end{equation}
where $\tilde{d}_{k,0}^{(1)}$ denotes the minimum access distance from user $k\in\mathcal{K}_0$ to $L-K+1$ BS antenna clusters which are uniformly distributed in the inscribed circle of Cell 0. $\mathbf{\tilde{H}}_{k,0}^{(1)}\in\mathbb{C}^{N\times N}$ denotes the corresponding small-scale fading matrix.
As $N\to\infty$, the empirical eigenvalue distribution of $\frac{1}{N}\mathbf{\tilde{H}}_{k,0}^{(1)}\left(\mathbf{\tilde{H}}_{k,0}^{(1)}\right)^\dag\sim \mathcal{W}_N\left(N,\frac{1}{N}\mathbf{I}_N\right)$ converges almost surely to the distribution given in (\ref{dis_da su}). By combining (\ref{Rklb}) and (\ref{dis_da su}), the asymptotic lower-bound of the per-antenna rate with the DA layout as $N\to\infty$ can be obtained as
\begin{equation}\label{Rk bd da}
    R_{k,lb}^{M-D}=\Phi\left(\frac{\frac{1}{K}\left(\tilde{d}_{k,0}^{(1)}\right)^{-\alpha}}{\frac{N_0}{P_t}+\frac{1}{L}\sum_{i=1}^6\sum_{l=1}^Ld_{k,l,i}^{-\alpha}} \right),
\end{equation}
where $\Phi(x)$ is defined in (\ref{Phi}).
With $L\gg K$, $\tilde{d}_{k,0}^{(1)}\ll 1$. The asymptotic lower-bound $R_{k,lb}^{M-D}$ can be then approximated by
\begin{align}\label{Rk bd da approx}
    R_{k,lb}^{M-D}\approx\log_2\left(\frac{\frac{1}{K}\left(\tilde{d}_{k,0}^{(1)}\right)^{-\alpha}}
    {\frac{N_0}{P_t}+\frac{1}{L}\sum_{i=1}^6\sum_{l=1}^Ld_{k,l,i}^{-\alpha}} \right)-\log_2e \mathop{\approx}\limits^{\textrm{for large }\tfrac{P_t}{N_0}} \log_2\left(\frac{\frac{1}{K}\left(\tilde{d}_{k,0}^{(1)}\right)^{-\alpha}}
    {\frac{1}{L}\sum_{i=1}^6\sum_{l=1}^Ld_{k,l,i}^{-\alpha}} \right)-\log_2e.
\end{align}

By combining  (\ref{Rk bd da approx}) and (\ref{Cav def}), the asymptotic lower-bound of the average per-antenna rate with the DA layout can be written as
\begin{align}\label{Rav da lb}
    \bar{R}_{lb}^{M-D}&= \mathbb{E}_{\rho_k,\tilde{d}_{k,0}^{(1)}}\left[\log_2\left(\frac{1}{K}\left(\tilde{d}_{k,0}^{(1)}\right)^{-\alpha}\right)\right]
    -\mathbb{E}_{\rho_k,\theta_k}\left[\log_2\left(\sum_{i=1}^6 \mathbb{E}_{d_{k,l,i}|\rho_k,\theta_k}\left[d_{k,l,i}^{-\alpha}|\rho_k,\theta_k\right]\right)\right]-\log_2e \nonumber \\
   &=2\int_0^1\int_0^{1+y}y\log_2\left(\frac{1}{K}x^{-\alpha}\right)f_{\tilde{d}_{k,0}^{(1)}|\rho_k}(x|y)dxdy+\Psi^D(\alpha),
\end{align}
where $f_{\tilde{d}_{k,0}^{(1)}|\rho_k}(x|y)$ denotes the conditional pdf of $\tilde{d}_{k,0}^{(1)}$ given user $k$'s position at $(\rho_k,\theta_k)$. Recall that $\tilde{d}_{k,0}^{(1)}$ is the minimum access distance from user $k$ to $L-K+1$ uniformly distributed antenna clusters in Cell 0. It can be easily obtained that
\begin{equation}\label{pdf_dmin bd}
    f_{\tilde{d}_{k,0}^{(1)}|\rho_k}(x|y)=(L-K+1)(1-F_{d_{k,l,0}|\rho_k}(x|y))^{L-K}f_{d_{k,l,0}|\rho_k}(x|y),
\end{equation}
where $F_{d_{k,l,0}|\rho_k}(x|y)$ and $f_{d_{k,l,0}|\rho_k}(x|y)$ are given in (\ref{cdf d0}) and (\ref{pdf d0}), respectively.
$\Psi^D(\alpha)$ denotes the sum of the last two items on the right-hand side of (\ref{Rav da lb}), which can be obtained as
\begin{align}\label{Psi da}
\Psi^D(\alpha)=-\frac{1}{\pi}\int_0^{2\pi}\int_0^1 y\log_2\left(\sum_{i=1}^6 \int_{\sqrt{y^2+4-4y\cos\left(z-\left(i\cdot \frac{\pi}{3}-\frac{\pi}{6}\right)\right)}-1}^{\sqrt{y^2+4-4y\cos\left(z-\left(i\cdot \frac{\pi}{3}-\frac{\pi}{6}\right)\right)}+1}
    x^{-\alpha}f_{d_{k,l,i}|\rho_k,\theta_k}(x|y,z) dx\right)dydz-\log_2e,
\end{align}
where $f_{d_{k,l,i}|\rho_k,\theta_k}(x|y,z) $ is given in (\ref{pdf d}). With the path-loss factor $\alpha=4$, for instance, we have $\Psi^D(4)\approx -3.054$.

\begin{figure}
\begin{center}
\includegraphics[width=0.6\textwidth]{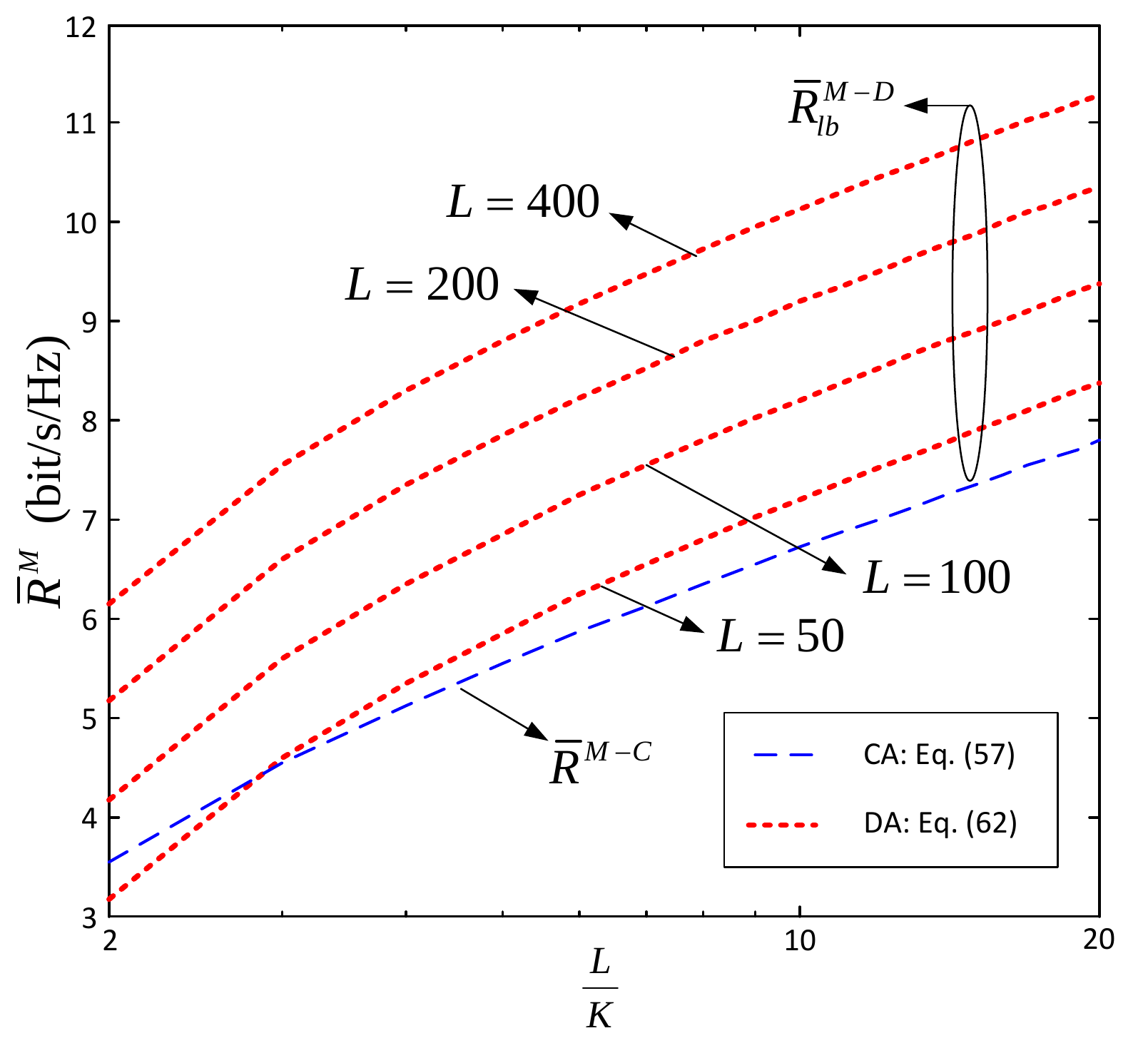}
\caption{Asymptotic average per-antenna rate with the CA layout $\bar{R}^{M-C}$ and the asymptotic lower-bound of the average per-antenna rate with the DA layout $\bar{R}_{lb}^{M-D}$ versus $\frac{L}{K}$. $\alpha=4$.}
\label{FIG_Cav_mc}
\end{center}
\end{figure}

Fig. \ref{FIG_Cav_mc} plots the asymptotic average per-antenna rate with the CA layout $\bar{R}^{M-C}$ and the asymptotic lower-bound of the average per-antenna rate with the DA layout $\bar{R}_{lb}^{M-D}$. We can see from Fig. \ref{FIG_Cav_mc} that both $\bar{R}^{M-C}$ and $\bar{R}_{lb}^{M-D}$ logarithmically increase with $\frac{L}{K}$. In contrast to $\bar{R}^{M-C}$ which is solely determined by $\frac{L}{K}$, $\bar{R}^{M-D}_{lb}$ can be further improved by increasing $L$. In fact, (\ref{Rav da lb}) has shown that with the DA layout, the asymptotic lower-bound $\bar{R}_{lb}^{M-D}$ is determined by the minimum access distance $\tilde{d}_{k,0}^{(1)}$, which decreases in the order of $(L-K+1)^{-1/2}$ as the number of BS antenna clusters $L$ increases according to (\ref{pdf_dmin bd}). As a result, $\bar{R}_{lb}^{M-D}$ scales in the order of $\log_2\frac{(L-K+1)^{\alpha/2}}{K}$, which can be much higher than $\bar{R}^{M-C}$ when $L$ is large.

\begin{figure*}
\centerline{\subfloat[]{\includegraphics[width=3.5in]{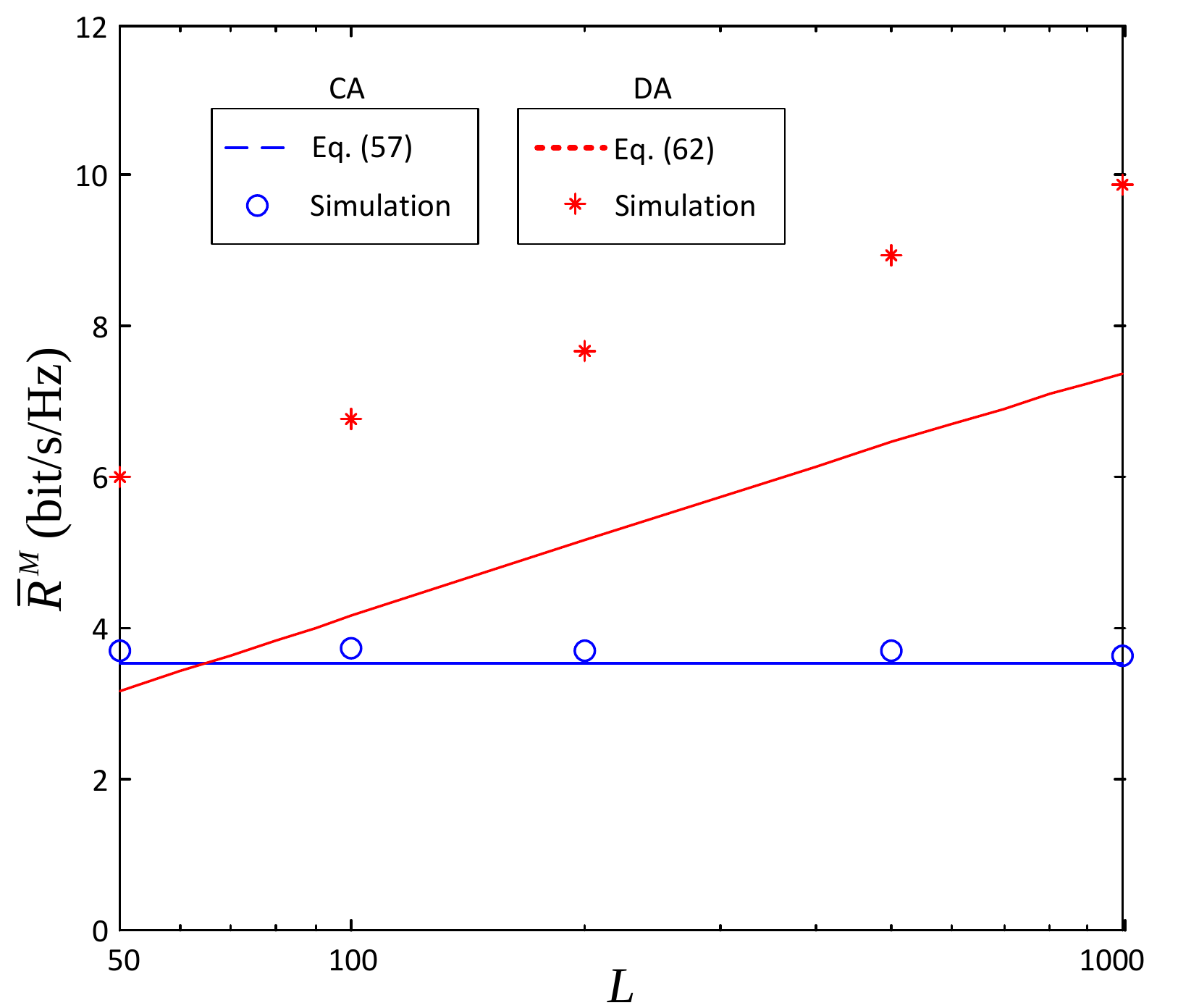}
\label{FIG_xi5}}\hfil
\subfloat[]{\includegraphics[width=3.5in]{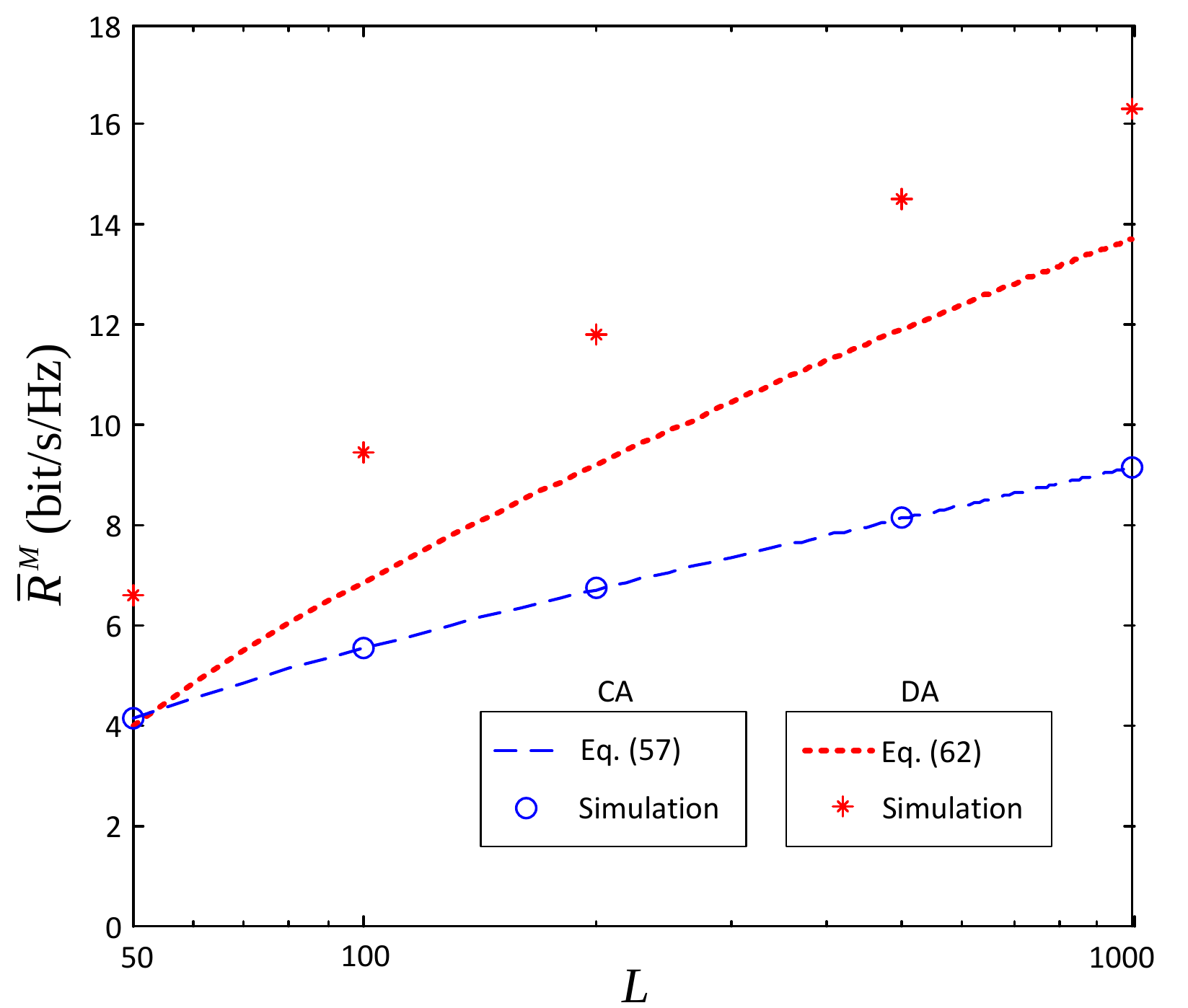}
\label{FIG_K50}}} \caption{Average per-antenna rate $\bar{R}^M$ versus the ratio $L$ of the number of BS antennas $M$ to the number of user antennas $N$ with (a) $\frac{L}{K}=2$ and (b) $K=20$. $\alpha=4$, $N=2$, $P_t/N_0=10$dB.}
\label{FIG_Cav}
\end{figure*}

The analysis is verified by the simulation results presented in Fig. \ref{FIG_Cav}. With the CA layout, the average per-antenna rate is obtained by averaging over 100 realizations of the users' positions. With the DA layout, it is further averaged over 100 realizations of the BS antenna topology. As we can see from Fig. \ref{FIG_xi5}, with the ratio $\tfrac{L}{K}$ fixed to be 2, the average per-antenna rate with the CA layout does not vary with $L$. Yet with the DA layout, similar to its asymptotic lower-bound $\bar{R}^{M-D}_{lb}$, the average per-antenna rate logarithmically increases with $L$ in the order of $(\tfrac{\alpha}{2}-1)\log_2L$, where the path-loss factor $\alpha>2$. In Fig. \ref{FIG_K50}, the number of users $K$ is fixed to be 20. In this case, it has been shown that $\bar{R}_{lb}^{M-D}$ and $\bar{R}^{M-C}$ scale in the orders of $\tfrac{\alpha}{2}\log_2 L$ and $\log_2 L$, respectively. As we can see from Fig. \ref{FIG_K50}, the average per-antenna rates with both CA and DA layouts logarithmically increase with $L$, and a much higher increasing rate is observed in the DA case. We can conclude from Figs. \ref{FIG_su_L} and \ref{FIG_Cav} that similar to the single-user case, the DA layout has a much higher average per-antenna rate than the CA layout when the number of BS antennas is large. The rate gains mainly come from the reduction of the minimum access distance, and become increasingly prominent as the number of BS antennas grows.

\begin{figure}
\begin{center}
\includegraphics[width=0.6\textwidth]{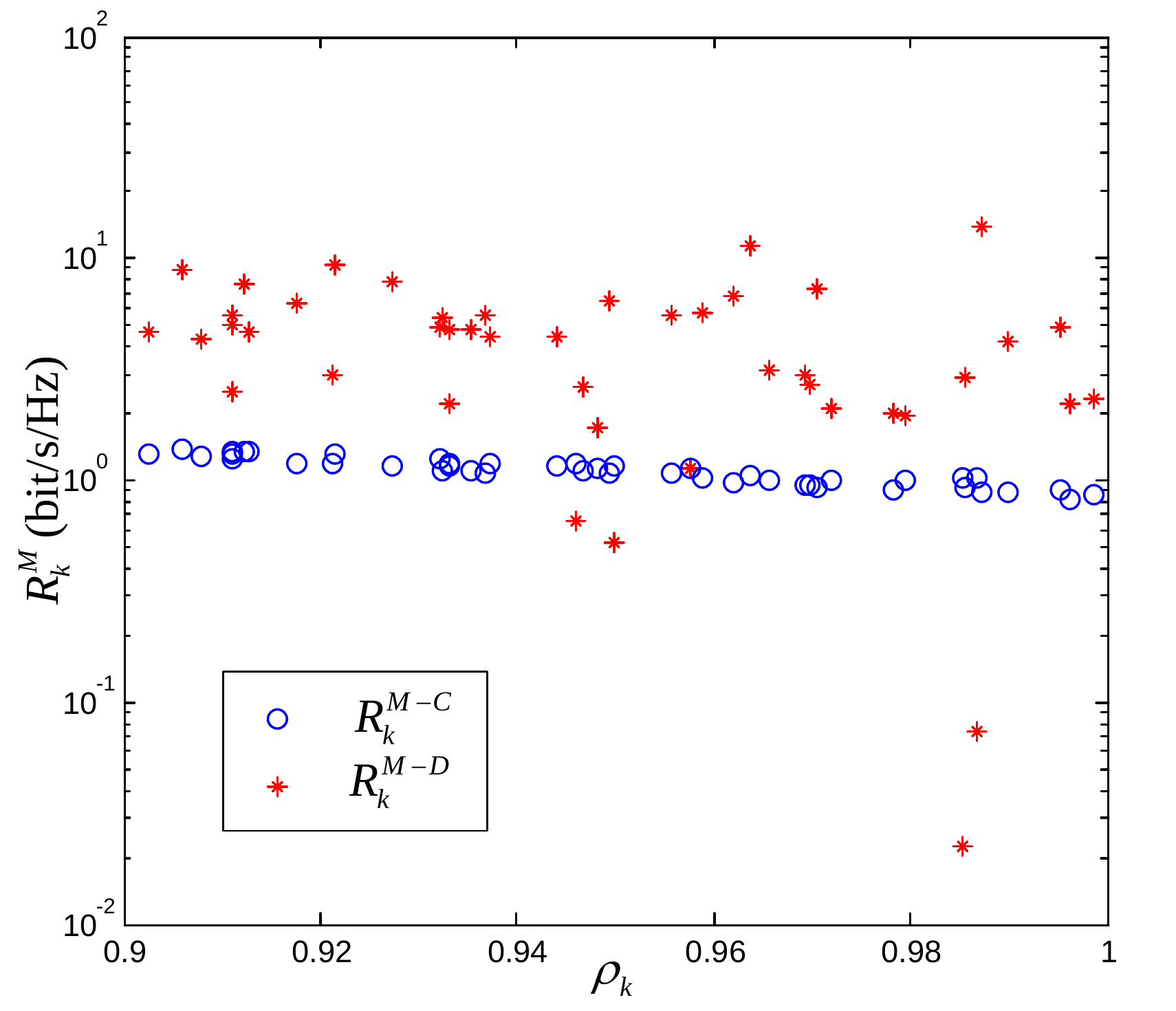}
\caption{Simulated per-antenna rate $R_k^{M}$ of user $k$ versus its radial coordinate $\rho_k$. $L=400$, $K=200$, $N=2$, $M=800$, $\alpha=4$, $P_t/N_0=10$dB.}
\label{FIG_Rktheta}
\end{center}
\end{figure}

Note that it comes with a caveat. Recall that it has been shown in Section IV-A that with the DA layout, the inter-cell interference level may be significantly enhanced at the cell edge, and becomes sensitive to the user's position. Fig. \ref{FIG_Rktheta} illustrates the corresponding per-antenna rate performance. We can clearly see from Fig. \ref{FIG_Rktheta} that compared to the CA layout, the per-antenna rate with the DA layout has a much larger variance. Intuitively, with a large amount of distributed BS antennas, the minimum access distance of each user is greatly reduced on average. Yet the chance that a cell-edge user is close to some BS antenna in the neighboring cells also becomes substantially higher. As a result, although most users can achieve better rate performance than that with the CA layout, a few ``unlucky'' ones may suffer from strong inter-cell interference due to their disadvantageous locations. As Fig. \ref{FIG_Rktheta} shows, with the DA layout, the per-antenna rate significantly varies with the user's position. Despite the improvement in the average rate performance, the rate difference among cell-edge users is greatly enlarged.

\section{Implications for Cellular Network Design}

So far we have shown that the rate scaling behavior of cellular networks closely depends on the BS antenna layout. The DA layout achieves a higher scaling order, and the rate gain over the CA layout continues to increase as more BS antennas are used. For the next-generation cellular networks where a large amount of BS antennas are expected to be deployed to meet the ever-increasing demand of high data rate, such a prominent rate gain may serve as a strong justification for the high implementation cost of distributed BS antennas.

\begin{figure}
\begin{center}
\includegraphics[width=0.4\textwidth]{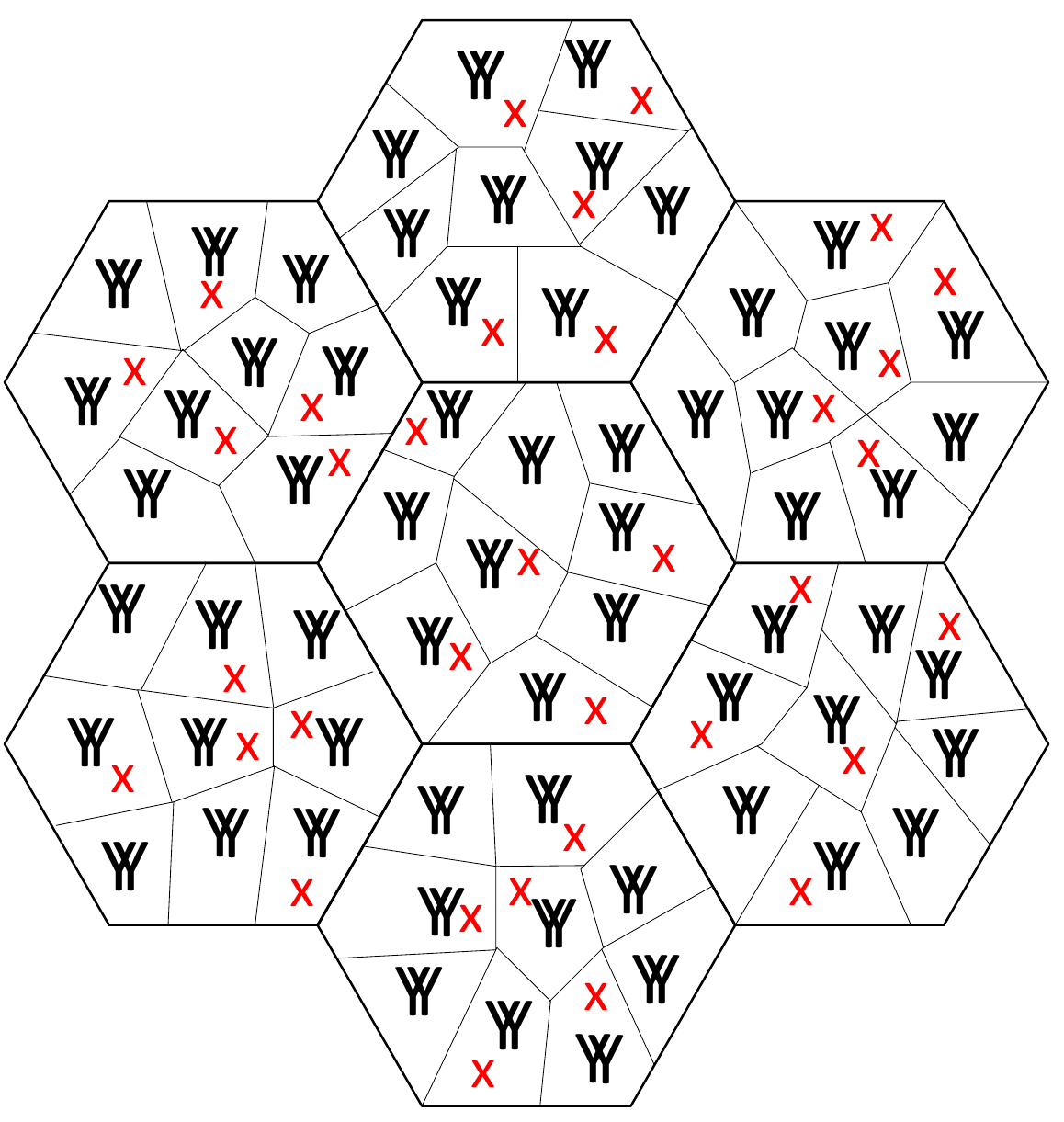}
\caption{Graphic illustration of an MIMO cellular network with small cells. Each conventional hexagonal cell is split into $L$ small cells with $N$ co-located BS antennas in each small cell. ``Y'' represents a BS antenna, and ``X'' represents a user.}
\label{FIG_SC}
\end{center}
\end{figure}

Note that in addition to employing more BS antennas in each cell, reducing the cell size is also a viable solution for improving the data rate. As Fig. \ref{FIG_SC} illustrates, a small-cell network is reminiscent of a cellular network with the DA layout, except that the BS antennas in different small cells transmit independently. Although the signal quality can be significantly enhanced by reducing the cell size, users may suffer from severe inter-cell interference, which greatly limits the rate performance.

Specifically, let us consider the small-cell network shown in Fig. \ref{FIG_SC}. For the sake of comparison, we assume that each cell is split into $L$ small cells with $N$ co-located BS antennas in each small cell. Similar to the DA layout, the BSs of $L$ small cells are supposed to be uniformly distributed in the inscribed circle of each hexagonal cell. As no coordination is adopted among small cells, each BS can serve at most one user if BD is adopted. The total number of users that are served by $L$ small cells remains to be $K$, and the transmit power for each user is $P_t/K$. For illustration, we focus on the average per-antenna rate of user $k$ at $(0,0)$. Appendix \ref{app sc} shows that as $N\to\infty$, an asymptotic lower-bound of the average per-antenna rate with small cells can be obtained as
\begin{equation}\label{Rav lb sc}
    \bar{R}_{lb}^{M-S}=\int_0^1 \Phi\left(\frac{\alpha-2}{2}\cdot \frac{x^{-2}}{K}\right) f_{d_{k,0}^{(1)}|\rho_k}(x|0) dx,
\end{equation}
where $\Phi(x)$ is given in (\ref{Phi}), and $f_{d_{k,0}^{(1)}|\rho_k}(x|0)$ denotes the pdf of the access distance from user $k$ at $(0,0)$ to its BS, which is given in (\ref{pdf dmin 0}).

\begin{figure}
\begin{center}
\includegraphics[width=0.6\textwidth]{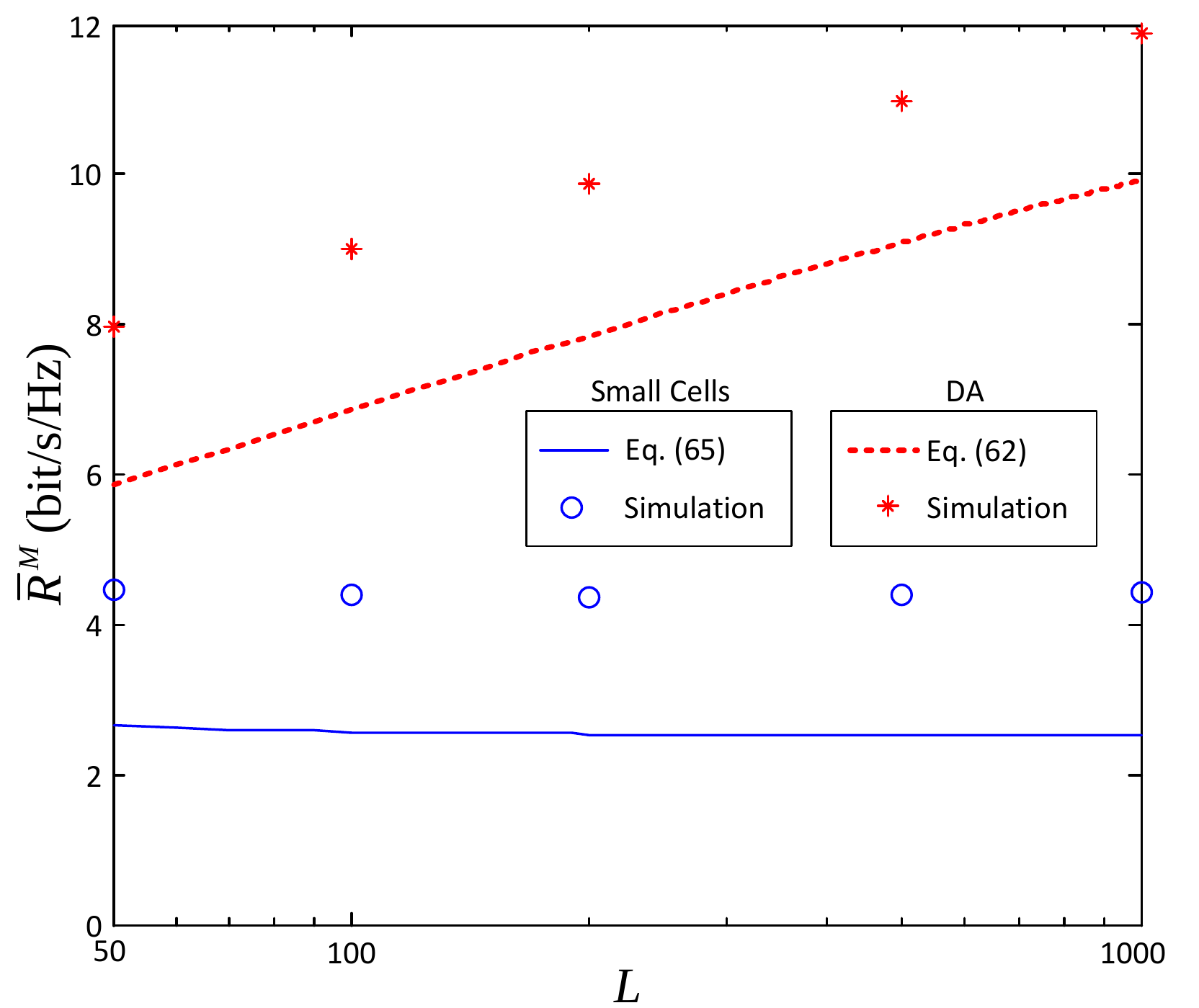}
\caption{Average per-antenna rate $\bar{R}^M$ versus the number of small cells $L$ in each cell. $\frac{L}{K}=5$, $\alpha=4$, $N=2$, $\frac{P_t}{N_0}=10$dB.}
\label{FIG_small_cell}
\end{center}
\end{figure}

Fig. \ref{FIG_small_cell} demonstrates the average per-antenna rate with small cells and its asymptotic lower-bound $\bar{R}_{lb}^{M-S}$. The average per-antenna rate with the DA layout and its asymptotic lower-bound $\bar{R}_{lb}^{M-D}$ are also plotted for comparison. It can be observed from Fig. \ref{FIG_small_cell} that similar to its asymptotic lower-bound $\bar{R}_{lb}^{M-S}$, the average per-antenna rate with small cells does not vary with $L$ when $\frac{L}{K}$ is fixed. Intuitively, as the number of small cells $L$ increases, the access distance $d_{k,0}^{(1)}$ from user $k$ to its BS decreases in the order of $L^{-1/2}$. As a result, we can see from (\ref{Rav lb sc}) that the asymptotic lower-bound $\bar{R}_{lb}^{M-S}$ scales in the order of $\log_2 \frac{L}{K}$, which is much smaller than that of $\bar{R}_{lb}^{M-D}$, i.e., $\log_2 \frac{(L-K+1)^{\alpha/2}}{K}$, where the path-loss factor $\alpha>2$. As Fig. \ref{FIG_small_cell} illustrates, the average per-antenna rate with small cells is significantly lower than that with the DA layout, and the rate gap is enlarged as the number of small cells $L$ increases due to different scaling orders. We can conclude from the comparison that coordination among distributed BS antennas is crucial for achieving the potential of high data rate.


\section{Conclusion}

In this paper, we present a comparative study on the ergodic rate performance of downlink MIMO cellular networks with the CA and DA layouts. By assuming that the number of BS antennas $M$ and the number of user antennas $N$ grow infinitely and $M/N\to L \gg 1$, the asymptotic average per-antenna capacity with the CA layout and an asymptotic lower-bound of the average per-antenna capacity with the DA layout in the single-user case are derived, which are shown to be logarithmically increasing with $L$, but in the orders of $\log_2 L$ and $\tfrac{\alpha}{2}\log_2 L$, respectively, where $\alpha>2$ is the path-loss factor. The analysis is further extended to a 1-tier MIMO cellular network with $K\gg 1$ users in each cell and BD adopted at each BS. With $M,N\to\infty$ and $M/N\to L\gg K$, the scaling orders of the asymptotic average per-antenna rate with the CA layout and the asymptotic lower-bound of the average per-antenna rate with the DA layout are found to be $\log_2 \frac{L}{K}$ and $\log_2 \frac{(L-K+1)^{\alpha/2}}{K}$, respectively. Simulation results verify that the average per-antenna rate with the DA layout scales with $L$ in the same order as its asymptotic lower-bound in both the single-user and multi-user cases. Substantial gains over the CA layout are observed when the ratio $L$ of the number of BS antennas to the number of user antennas is large, which are mainly attributed to the reduction of minimum access distance.

Despite better average rate performance, the inter-cell interference with the DA layout is shown to be sensitive to the user's position at the cell edge, leading to a large rate difference among cell-edge users. To achieve a uniform rate across the cell, proper transmit power allocation should be performed at each BS. With a large number of distributed BS antennas, how to allocate the transmit power to maintain a constant SINR for all the users is a challenging issue, which deserves much attention in the future study.

\appendices

\section{Derivation of (\ref{intercell})} \label{app inter}

Let $q_{l,t}^{inter}$ denote the entry of $\mathbf{Q}_{k}^{inter}$ at the $l$-th row and $t$-th column. It can be written as
\begin{equation}\label{app inter lt}
    q_{l,t}^{inter}{=}{}\sum_{i=1}^6 \sum_{j\in \mathcal{K}_i}\frac{P_t}{K} \mathbb{E}\left[\sum_{n=1}^N \mathbf{g}_{l}^{k,\mathcal{B}_i}\mathbf{w}_n^j\left(\mathbf{g}_{t}^{k,\mathcal{B}_i}\mathbf{w}_n^j\right)^\dag\right],
\end{equation}
where $\mathbf{g}_{l}^{k,\mathcal{B}_i}\in \mathbb{C}^{1\times M}$ denotes the $l$-th row vector of $\mathbf{G}_{k,\mathcal{B}_i}$ and $\mathbf{w}_n^j\in\mathbb{C}^{M\times 1}$ denotes the $n$-th column vector of $\mathbf{W}_j$. Note that the precoding matrix $\mathbf{W}_j$ of user $j\in \mathcal{K}_i$ is independent of the channel gain matrix $\mathbf{G}_{k,\mathcal{B}_i}$ from BS antennas in Cell $i$ to user $k\in\mathcal{K}_0$, $i=1,\cdots,6$. Therefore we have
\begin{equation}\label{app_2}
q_{l,t}^{inter}=\left\{\begin{array}{c} P_t\sum_{i=1}^{6} \sum_{m\in\mathcal{B}_i}|\gamma_{k,m}|^2\sum_{n=1}^N\frac{1}{K}\sum_{j\in\mathcal{K}_i}\mathbb{E}\left[|w_{m,n}^j|^2\right]\\ {0} \end{array}\right. \begin{array}{cc} {} & {l=t} \\ {} & {l\ne t,} \end{array}
\end{equation}
where $w_{m,n}^j$ denotes the entry of $\mathbf{W}_j$ at the $m$-th row and $n$-th column. As the number of users $K\to\infty$, we have $\frac{1}{K}\sum_{j\in\mathcal{K}_i} \mathbb{E}\left[|w_{m,n}^j|^2\right]\to\frac{1}{MN}$. The covariance $\mathbf{Q}_k^{inter}$ can be then obtained as (\ref{intercell}).

\section{Derivation of (\ref{pdf d})} \label{app pdf d}

In each cell, BS antenna clusters are uniformly distributed over the inscribed circle with radius 1 centered at $O_i=(\rho_i,\theta_i)$, where $O_0=(0,0)$ and $O_i=(2,i\cdot\tfrac{\pi}{3}-\tfrac{\pi}{6})$ for $i=1,\cdots,6$. For user $k\in\mathcal{K}_0$ at $(\rho_k,\theta_k)$, the conditional probability distribution function (pdf) of the distance $d_{k,l,i}$ from user $k$ to BS antenna cluster $l$ in Cell $i$, $l=1,\cdots,L$, $i=1,\cdots,6$, is given by
\begin{equation}\label{app pdf}
    f_{d_{k,l,i}|\rho_k,\theta_k}(x|y,z)=dF_{d_{k,l,i}|\rho_k,\theta_k}(x|y,z)/dx,
\end{equation}

\begin{figure}
\begin{center}
\includegraphics[width=0.6\textwidth]{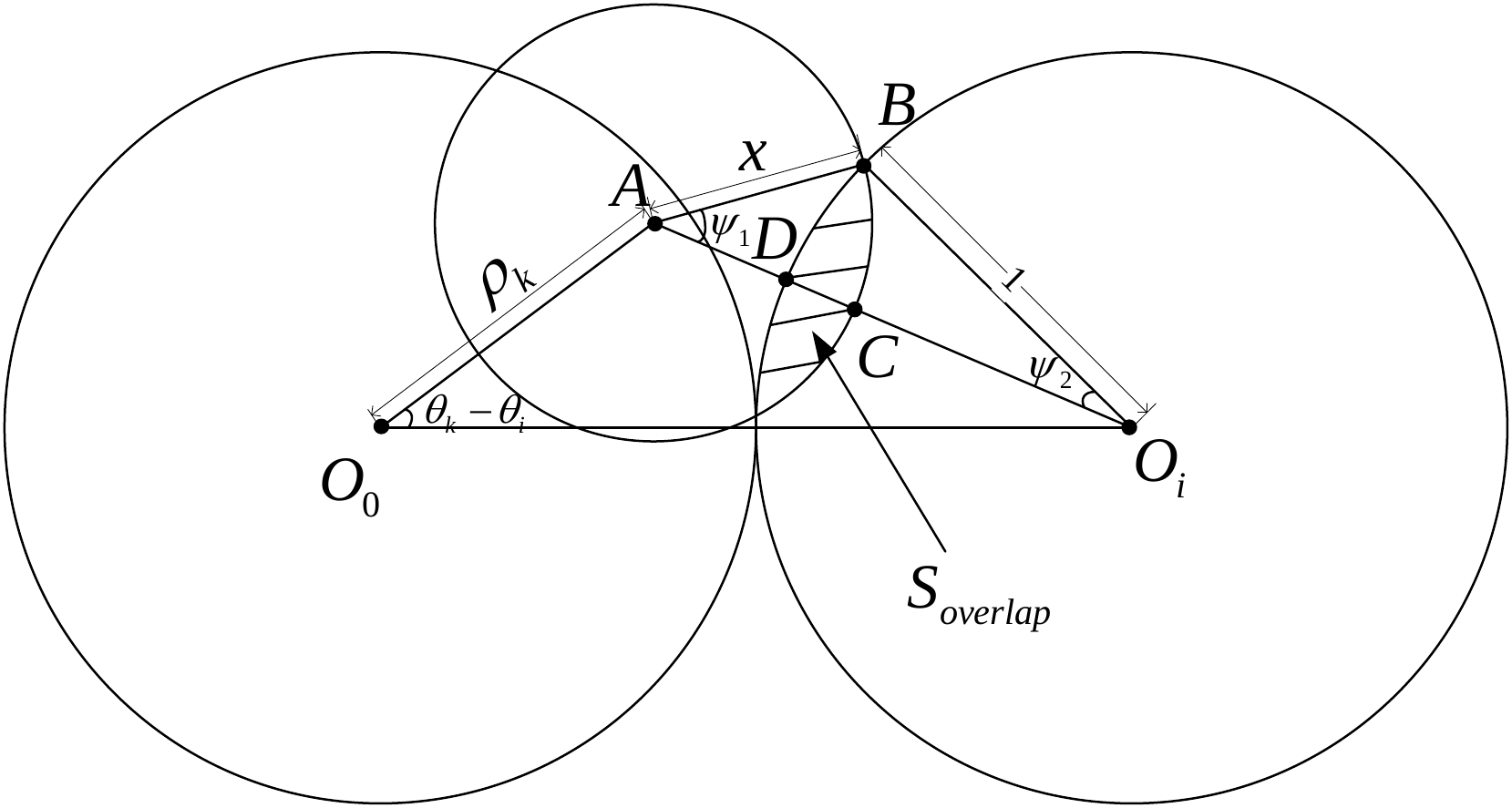}
\caption{Graphic illustration of $S_{overlap}$.}
\label{FIG_overlap}
\end{center}
\end{figure}

\noindent where $F_{d_{k,l,i}|\rho_k,\theta_k}(x|y,z)$ is the conditional cumulative density function (cdf) of $d_{k,l,i}$ given the position of user $k$, which is given by
\begin{equation}\label{app4 cdf}
    F_{d_{k,l,i}|\rho_k,\theta_k}(x|y,z)=\frac{S_{overlap}}{\pi },
\end{equation}
where $S_{overlap}$ is the intersection area of the circle with center $O_i$ and radius 1 and the circle with center $A$ and radius $x$, as shown in Fig. \ref{FIG_overlap}. It can be obtained that
\begin{equation}\label{app s overlap}
    S_{overlap}=2(S_{ABC}+S_{O_iBD}-S_{\triangle ABO_i}),
\end{equation}
where $S_{ABC}$ and $S_{O_iBD}$ denote the areas of circular sectors $ABC$ and $O_iBD$, respectively, which are given by
\begin{equation}\label{app s abc}
    S_{ABC}=\frac{\psi_1}{2\pi}\cdot \pi x^2,
\end{equation}
and
\begin{equation}\label{app s oibd}
    S_{O_iBD}=\frac{\psi_2}{2\pi}\cdot \pi x^2,
\end{equation}
with
\begin{equation}\label{app cos psi1}
{\psi _1} =\arccos \frac{{{x^2} + \left(y^2+4-4y\cos\left(z-\left(i\cdot \frac{\pi}{3}-\frac{\pi}{6}\right)\right)\right) - {1}}}{2x\sqrt{y^2+4-4y\cos\left(z-\left(i\cdot \frac{\pi}{3}-\frac{\pi}{6}\right)\right)}},
\end{equation}
and
\begin{equation}\label{app cos psi2}
{\psi _2} = \arccos \frac{{1 + \left(y^2+4-4y\cos\left(z-\left(i\cdot \frac{\pi}{3}-\frac{\pi}{6}\right)\right)\right) - {x^2}}}{2\sqrt{y^2+4-4y\cos\left(z-\left(i\cdot \frac{\pi}{3}-\frac{\pi}{6}\right)\right)}},
\end{equation}
if
\begin{equation}\label{app x}
    \sqrt{y^2+4-4y\cos\left(z-\left(i\cdot \frac{\pi}{3}-\frac{\pi}{6}\right)\right)}-1\leq x\leq \sqrt{y^2+4-4y\cos\left(z-\left(i\cdot \frac{\pi}{3}-\frac{\pi}{6}\right)\right)}+1,
\end{equation}
and otherwise $\psi_1=\psi_2=0$.
$S_{\triangle ABO_i}$ is the area of $\triangle ABO_i$, which is given by
\begin{equation}\label{app s aboi}
    S_{\triangle ABO_i}=\frac{1}{2}x\sqrt{y^2+4-4y\cos\left(z-\left(i\cdot \frac{\pi}{3}-\frac{\pi}{6}\right)\right)}\sin\psi_1.
\end{equation}
(\ref{pdf d}) can be then obtained by combining (\ref{app pdf}-\ref{app s aboi}).

\section{Derivation of (\ref{Capacity su da 1})} \label{app su}
According to (\ref{Ck su}), the per-antenna capacity is lower-bounded by
\begin{equation}\label{app1 epa}
    R_k^{S}>\frac{1}{N}\mathbb{E}_{\mathbf{H}_{k,\mathcal{B}_0}}\left[\log_2\det\left(\mathbf{I}_N+\frac{1}{N}\frac{\bar{P}_k}{N_0} \mathbf{G}_{k,\mathcal{B}_0}\mathbf{G}_{k,\mathcal{B}_0}^\dag\right)\right],
\end{equation}
where the right-hand side of (\ref{app1 epa}) is obtained by applying equal power allocation over $N$ sub-channels. With the DA layout, the channel gain matrix $\mathbf{G}_{k,\mathcal{B}_0}^D$ can be written as
\begin{equation}\label{app su G}
    \mathbf{G}_{k,\mathcal{B}_0}^D=\left[d_{k,1,0}^{-\alpha/2}\mathbf{H}_{k,1,0},\cdots, d_{k,L,0}^{-\alpha/2}\mathbf{H}_{k,L,0}\right],
\end{equation}
where $d_{k,l,0}$ and $\mathbf{H}_{k,l,0}\in \mathbb{C}^{N\times N}$ denote the access distance from user $k$ to BS antenna cluster $l$ in Cell $0$ and the corresponding small-scale fading matrix, respectively, $l=1,\cdots, L$.

We can further obtain from (\ref{app su G}) that
\begin{equation}\label{app su ineq}
    \mathbf{G}_{k,\mathcal{B}_0}^D\left(\mathbf{G}_{k,\mathcal{B}_0}^D\right)^{\dag}
    =\sum_{l=1}^L d_{k,l,0}^{-\alpha}\mathbf{H}_{k,l,0}\mathbf{H}_{k,l,0}^{\dag}
    =\sum_{l=1}^L\left(d_{k,0}^{(l)}\right)^{-\alpha}\mathbf{H}_{k,0}^{(l)}\left(\mathbf{H}_{k,0}^{(l)}\right)^{\dag},
\end{equation}
where $d_{k,0}^{(l)}$ and $\mathbf{H}_{k,0}^{(l)}$ denote the access distance between user $k$ and the $l$-th closest BS antenna cluster in Cell 0 and the corresponding small-scale fading matrix, respectively, $l=1,\cdots,L$. Note that for $N\times N$ positive semi-definite Hermitian matrices $\mathbf{A}$ and $\mathbf{B}$, we have
\begin{equation}\label{app mdt}
\det(\mathbf{A}+\mathbf{B})^{\frac{1}{N}}\geq \det(\mathbf{A})^{\frac{1}{N}}+\det(\mathbf{B})^{\frac{1}{N}},
\end{equation}
according to Minkowski's determinant theorem \cite{Marcus1964}, where the equality holds when $\mathbf{A}=c\mathbf{B}$. For positive definite Hermitian matrices $\mathbf{A}$ and $\mathbf{B}$, we further have
\begin{equation}\label{app det}
\det(\mathbf{A}+\mathbf{B})>\det(\mathbf{A})+\det(\mathbf{B}).
\end{equation}
As $\left(d_{k,0}^{(l)}\right)^{-\alpha}\mathbf{H}_{k,0}^{(l)}\left(\mathbf{H}_{k,0}^{(l)}\right)^{\dag}$ is a positive definite Hermitian matrix, we then have
\begin{equation}\label{app su det}
    \det\left(\mathbf{I}_N+\frac{1}{N} \frac{\bar{P}_k}{N_0}\mathbf{G}_{k,\mathcal{B}_0}^D\left(\mathbf{G}_{k,\mathcal{B}_0}^D\right)^\dag\right) >\det\left(\mathbf{I}_N+\frac{1}{N}\frac{\bar{P}_k}{N_0}\left(d_{k,0}^{(1)}\right)^{-\alpha}\mathbf{H}_{k,0}^{(1)}\left(\mathbf{H}_{k,0}^{(1)}\right)^{\dag}\right).
\end{equation}
(\ref{Capacity su da 1}) can be then obtained by combining (\ref{app1 epa}) and (\ref{app su det}).

\section{Derivation of (\ref{Rklb})}\label{app bd}

With a large number of BS antenna clusters $L$, each user $k\in\mathcal{K}_0$ is close to some antenna cluster $l_k^*$, such that the large-scale fading coefficient $d_{k,l_k^*,0}^{-\alpha}\gg d_{k,l,0}^{-\alpha}$ if $l\ne l_k^*$. The normalized large-scale fading matrix can be then approximated by
\begin{equation}\label{B BD}
    \mathbf{B}_{k,\mathcal{B}_0}\approx \sqrt{\frac{1}{N}}\left[\mathbf{0}_{N\times N(l_k^*-1)},\mathbf{1}_{N\times N}, \mathbf{0}_{N\times N(L-l_k^*)}\right],
\end{equation}
according to (\ref{beta def}). Moreover, with $L\gg K$, the probability that user $j_1$ and user $j_2$ are close to the same BS antenna cluster is low, i.e., $\textrm{Pr}\{l_{j_1}^*=l_{j_2}^*|j_1\ne j_2\}\approx 0$. Denote $\mathcal{S}_k=\{1,\cdots,L\}-\{l_j^*\}_{j\in\mathcal{K}_0,j\ne k}$. As both $L$ BS antenna clusters and $K$ users are uniformly distributed, we can conclude that $\mathcal{S}_k$ is composed by $L-K+1$ uniformly distributed BS antenna clusters.
By combining (\ref{tildeG def}), (\ref{X}-\ref{X_SVD}) and (\ref{B BD}) , $\mathbf{\hat{V}}_{k,\mathcal{B}_0}^{(0)}$ can be written as
\begin{equation}\label{Vk0}
    \mathbf{\hat{V}}_{k,\mathcal{B}_0}^{(0)}\approx\left[\mathbf{E}_{l_1},\cdots, \mathbf{E}_{l_{L-K+1}}\right],
\end{equation}
where the $t$-th sub-matrix $\mathbf{E}_{l_t}\in \mathbb{C}^{M\times N}$ is given by
\begin{equation}
    \mathbf{E}_{l_t}=\left[\mathbf{0}_{N\times N(l_t-1)},\mathbf{I}_N,\mathbf{0}_{N\times N(L-l_t)}\right]^T,
\end{equation}
$l_t\in\mathcal{S}_k$, $t=1,\cdots,L-K+1$.

According to (\ref{Rk bd}), the per-antenna rate of user $k$ with BD is lower-bounded by
\begin{align}\label{app bd Rk1}
     R_k^{M}>\frac{1}{N}\mathbb{E}_{\mathbf{H}_{k,\mathcal{B}_0}}\left[\log_2\det\left(\mathbf{I}_N+\frac{\tilde{\mu}_k}{N}\mathbf{\tilde{X}}_{k,\mathcal{B}_0}\mathbf{\tilde{X}}_{k,\mathcal{B}_0}^\dag\right)\right],
\end{align}
where the right-hand side of (\ref{app bd Rk1}) is obtained by applying equal power allocation over $N$ sub-channels. Note that $\mathbf{\tilde{X}}_{k,\mathcal{B}_0}=\mathbf{\tilde{G}}_{k,\mathcal{B}_0}\mathbf{\hat{V}}_{k,\mathcal{B}_0}^{(0)}$. By combining (\ref{sinr}-\ref{Pk int def}), (\ref{inter da}) and (\ref{app bd Rk1}), the lower-bound of the per-antenna rate with the DA layout can be further written as
\begin{equation}\label{app bd Rk2} R_k^{M-D}>\frac{1}{N}\mathbb{E}_{\mathbf{H}_{k,\mathcal{B}_0}}\left[\log_2\det\left(\mathbf{I}_N+\frac{1}{N}\cdot\frac{\tfrac{1}{K}}{\frac{N_0}{P_t}+\frac{1}{L}\sum_{i=1}^6\sum_{l=1}^L d_{k,l,i}^{-\alpha}}\mathbf{G}_{k,\mathcal{B}_0}\mathbf{\hat{V}}_{k,\mathcal{B}_0}^{(0)}\left(\mathbf{G}_{k,\mathcal{B}_0}\mathbf{\hat{V}}_{k,\mathcal{B}_0}^{(0)}\right)^\dag\right)\right].
\end{equation}
According to (\ref{Vk0}), when $L\gg K$, $\mathbf{G}_{k,\mathcal{B}_0}\mathbf{\hat{V}}_{k,\mathcal{B}_0}^{(0)}$ can be approximated by
\begin{equation}\label{app bd Gk}
    \mathbf{G}_{k,\mathcal{B}_0}\mathbf{\hat{V}}_{k,\mathcal{B}_0}^{(0)}
    \approx \left[d_{k,l_1,0}^{-\alpha/2}\mathbf{H}_{k,l_1,0},\cdots,d_{k,l_{L-K+1},0}^{-\alpha/2}\mathbf{H}_{k,l_{L-K+1},0}\right],
\end{equation}
where $d_{k,l_t,0}$ and $\mathbf{H}_{k,l_t,0}$ denote the access distance from user $k$ to BS antenna cluster $l_t$ in Cell $0$ and the corresponding small-scale fading matrix, respectively, $l_t\in \mathcal{S}_k$, $t=1,\cdots,L-K+1$.
We can further obtain from (\ref{app bd Gk}) that
\begin{equation}\label{app bd Gk ineq}
    \mathbf{G}_{k,\mathcal{B}_0}\mathbf{\hat{V}}_{k,\mathcal{B}_0}^{(0)}\left(\mathbf{{G}}_{k,\mathcal{B}_0}\mathbf{\hat{V}}_{k,\mathcal{B}_0}^{(0)}\right)^\dag
    =\sum_{l_t\in \mathcal{S}_k} d_{k,l_t,0}^{-\alpha}\mathbf{H}_{k,l_t,0}\mathbf{H}_{k,l_t,0}^\dag
    =\sum_{l=1}^{L-K+1}\left(\tilde{d}_{k,0}^{(l)}\right)^{-\alpha} \mathbf{\tilde{H}}_{k,0}^{(l)}\left(\mathbf{\tilde{H}}_{k,0}^{(l)}\right)^\dag,
\end{equation}
where $\tilde{d}_{k,0}^{(l)}$ and $\mathbf{\tilde{H}}_{k,0}^{(l)}$ denote the access distance between user $k$ and the $l$-th closest BS antenna cluster in $\mathcal{S}_k$ and the corresponding small-scale fading matrix, $l=1,\cdots,L-K+1$. Note that $\left(\tilde{d}_{k,0}^{(l)}\right)^{-\alpha} \mathbf{\tilde{H}}_{k,0}^{(l)}\left(\mathbf{\tilde{H}}_{k,0}^{(l)}\right)^\dag$ is a positive definite Hermitian matrix. We then have
\begin{equation}\label{app bd det}
    \det\left(\mathbf{I}_N+ \mathbf{G}_{k,\mathcal{B}_0}\mathbf{\hat{V}}_{k,\mathcal{B}_0}^{(0)}\left(\mathbf{{G}}_{k,\mathcal{B}_0}\mathbf{\hat{V}}_{k,\mathcal{B}_0}^{(0)}\right)^\dag\right)>
    \det\left(\mathbf{I}_N+\left(\tilde{d}_{k,0}^{(1)}\right)^{-\alpha} \mathbf{\tilde{H}}_{k,0}^{(1)}\left(\mathbf{\tilde{H}}_{k,0}^{(1)}\right)^\dag\right),
\end{equation}
according to (\ref{app det}). (\ref{Rklb}) can be then obtained by combining (\ref{app bd Rk2}) and (\ref{app bd det}).

\section{Derivation of (\ref{Rav lb sc})} \label{app sc}

By following a similar derivation as Appendix \ref{app inter}, we can obtain the covariance matrix of inter-cell interference of user $k$ with small cells as
\begin{equation}\label{inter sc}
 \mathbf{Q}_k^{inter,S}=\frac{P_t}{K} \left(\sum_{j\in\mathcal{K}_0,j\ne k} d_{k,l_j^*,0}^{-\alpha}+\sum_{i=1}^6 \sum_{j\in\mathcal{K}_i} d_{k,l_j^*,i}^{-\alpha} \right)\mathbf{I}_N,
\end{equation}
where $l_j^*$ denotes the BS that serves user $j$, $j\in\mathcal{K}_i$, $i=0,\cdots,6$. By combining (\ref{inter sc}) and (\ref{Capacity su da 1}), the per-antenna rate of user $k$ with small cells is lower-bounded by
\begin{equation}\label{rk lb sc}
     R_{k,lb}^{M-S}=\frac{1}{N}\mathbb{E}_{\mathbf{H}_{k,0}^{(1)}}\left[\log_2 \det \left(\mathbf{I}_N+\frac{\frac{1}{K}\left(d_{k,0}^{(1)}\right)^{-\alpha}\cdot\frac{1}{N}\mathbf{H}_{k,0}^{(1)}\left(\mathbf{H}_{k,0}^{(1)}\right)^\dag}{\frac{N_0}{P_t}+ \frac{1}{K}\left(\sum_{j\in\mathcal{K}_0,j\ne k} d_{k,l_j^*,0}^{-\alpha}+\sum_{i=1}^6 \sum_{j\in\mathcal{K}_i} d_{k,l_j^*,i}^{-\alpha}\right)} \right)\right],
\end{equation}
where $d_{k,0}^{(1)}$ and $\mathbf{H}^{(1)}_{k,0}\in \mathbb{C}^{N\times N}$ denote the access distance from user $k$ to its BS and the corresponding small-scale fading matrix, respectively. As $N\to\infty$, the asymptotic lower-bound of the per-antenna rate of user $k$ can be obtained as
\begin{align}\label{rk lb sc 1}
    R_{k,lb}^{M-S}=\Phi\left(\frac{\frac{1}{K}\left(d_{k,0}^{(1)}\right)^{-\alpha}}{\frac{N_0}{P_t}+ \frac{1}{K} \left(\sum_{j\in\mathcal{K}_0,j\ne k} d_{k,l_j^*,0}^{-\alpha}+\sum_{i=1}^6 \sum_{j\in\mathcal{K}_i} d_{k,l_j^*,i}^{-\alpha}\right)}\right),
\end{align}
where $\Phi(x)$ is defined in (\ref{Phi}). By combining (\ref{rk lb sc 1}) and (\ref{Cav def}), the asymptotic lower-bound of the average per-antenna rate with small cells can be obtained as
\begin{equation}\label{Rav lb sc 1}
    \bar{R}^{M-S}>\bar{R}^{M-S}_{lb}=\mathbb{E}_{d_{k,0}^{(1)}}\left[ \Phi\left(\frac{\frac{1}{K}\left(d_{k,0}^{(1)}\right)^{-\alpha}}
    {\frac{N_0}{P_t}+\mathbb{E}_{j\in\mathcal{K}_0,j\ne k}\left[d_{k,l_j^*,0}^{-\alpha}\right]+\sum_{i=1}^6\mathbb{E}_{j\in\mathcal{K}_i}\left[d_{k,l_j^*,i}^{-\alpha}\right]}\right)\right].
\end{equation}

For user $k$ at $(0,0)$, the pdfs of its distances $d_{k,l,i}$ to the BS of small cell $l$ in Cell $i$, $i=0,\cdots,6$, can be easily obtained from (\ref{pdf d0}-\ref{pdf d range}) as
\begin{equation}\label{pdf d0 center}
   {f_{{d_{k,l,0}}|{\rho_k}}}(x|0) = \left\{ {\begin{array}{*{20}{c}}
   {2x} & {\textrm{if } 0 \leq x \leq 1}  \\
   0 & {\textrm{otherwise},}  \\
\end{array} } \right.
\end{equation}
and
\begin{equation}\label{pdf d center}
   {f_{{d_{k,l,i}}|{\rho_k},{\theta_k}}}(x|0,0) = \left\{ {\begin{array}{*{20}{c}}
   {\frac{2x}{\pi}\arccos \frac{{{x^2} + 3}}{{4 x}}} & {\textrm{if } 1 \leq x \leq 3}  \\
   0 & {\textrm{otherwise},}  \\
\end{array} } \right.
\end{equation}
$i=1,\cdots,6$, respectively. Therefore, we have
\begin{equation}\label{inter sc mean}
    \sum_{i=1}^6 \mathbb{E}_{j\in\mathcal{K}_i}\left[d_{k,l_j^*,i}^{-\alpha}\right]=6\Upsilon(\alpha),
\end{equation}
where $\Upsilon(\alpha)=\frac{2}{\pi}\int_1^3 x^{1-\alpha}\arccos\frac{x^2+3}{4x} dx$. With $\alpha=4$, for instance, we have $\Upsilon(4)=\frac{1}{9}$. For $j\in\mathcal{K}_0$, note that $d_{k,l_j^*,0}\geq d_{k,0}^{(1)}$ if $j\ne k$. The pdf of $d_{k,l_j^*,0}$ for user $k$ at $(0,0)$ and $j\in\mathcal{K}_0$, $j\ne k$, can be then obtained as
\begin{equation}\label{pdf_d_1}
f_{d_{k,l_j^*,0}|\rho_k}(x|0)=\left\{ {\begin{array}{*{20}{c}}
{\frac{2x}{1-\left(d_{k,0}^{(1)}\right)^{2}}} & \textrm{if } {d_{k,0}^{(1)}\leq x\leq1} \\
0 & {\textrm{otherwise}.}
\end{array} } \right.
\end{equation}
We then have
\begin{equation}\label{inter sc mean 0}
    \mathbb{E}_{j\in\mathcal{K}_0,j\ne k}\left[d_{k,l_j^*,0}^{-\alpha}\right]=\frac{2}{\alpha-2}\cdot \frac{\left(d_{k,0}^{(1)}\right)^{2-\alpha}-1}{1-\left(d_{k,0}^{(1)}\right)^2}.
\end{equation}
By combining (\ref{Rav lb sc 1}), (\ref{inter sc mean}) and (\ref{inter sc mean 0}), the asymptotic lower-bound of the average per-antenna rate of user $k$ at $(0,0)$ can be written as
\begin{equation}\label{Rav lb sc 2}
\bar{R}^{M-S}_{lb}=\int_0^1 \Phi\left(\frac{\frac{1}{K}}
    {\left(d_{k,0}^{(1)}\right)^{\alpha}\left(\frac{N_0}{P_t}+6\Upsilon(\alpha)\right)+\frac{2}{\alpha-2} \frac{\left(d_{k,0}^{(1)}\right)^{2}-\left(d_{k,0}^{(1)}\right)^{\alpha}}{1-\left(d_{k,0}^{(1)}\right)^2}}\right)\cdot f_{d_{k,0}^{(1)}|\rho_k}(x|0) dx,
\end{equation}
where the pdf of the access distance from user $k$ at $(0,0)$ to its BS can be obtained by combining (\ref{pdf d0 center}) and (\ref{pdf_dmin}) as
\begin{equation}\label{pdf dmin 0}
    f_{d_{k,0}^{(1)}|\rho_k}(x|0)=\left\{ {\begin{array}{*{20}{c}}
        2Lx(1-x^2)^{L-1} & {\textrm{if } 0 \leq x \leq 1} \\
        0 & {\textrm{otherwise}.}
\end{array} } \right.
\end{equation}
With $L\gg 1$, $d_{k,0}^{(1)}\ll 1$. We then have
\begin{equation}\label{Rav lb sc approx}
   \Phi\left(\frac{\frac{1}{K}}
    {\left(d_{k,0}^{(1)}\right)^{\alpha}\left(\frac{N_0}{P_t}+6\Upsilon(\alpha)\right)+\frac{2}{\alpha-2} \frac{\left(d_{k,0}^{(1)}\right)^{2}-\left(d_{k,0}^{(1)}\right)^{\alpha}}{1-\left(d_{k,0}^{(1)}\right)^2}}\right)
    \approx \Phi\left(\frac{\alpha-2}{2K}\left(d_{k,0}^{(1)}\right)^{-2}\right).
\end{equation}
(\ref{Rav lb sc}) can be then obtained by combining (\ref{Rav lb sc 2}) and (\ref{Rav lb sc approx}).

\vspace{3mm}

\bibliography{ieeeabrv,das_bib}

\begin{thebibliography}{10}
\providecommand{\url}[1]{#1}
\csname url@samestyle\endcsname
\providecommand{\newblock}{\relax}
\providecommand{\bibinfo}[2]{#2}
\providecommand{\BIBentrySTDinterwordspacing}{\spaceskip=0pt\relax}
\providecommand{\BIBentryALTinterwordstretchfactor}{4}
\providecommand{\BIBentryALTinterwordspacing}{\spaceskip=\fontdimen2\font plus
\BIBentryALTinterwordstretchfactor\fontdimen3\font minus
  \fontdimen4\font\relax}
\providecommand{\BIBforeignlanguage}[2]{{%
\expandafter\ifx\csname l@#1\endcsname\relax
\typeout{** WARNING: IEEEtran.bst: No hyphenation pattern has been}%
\typeout{** loaded for the language `#1'. Using the pattern for}%
\typeout{** the default language instead.}%
\else
\language=\csname l@#1\endcsname
\fi
#2}}
\providecommand{\BIBdecl}{\relax}
\BIBdecl

\bibitem{Marzetta2010}
T.~Marzetta, ``Noncooperative cellular wireless with unlimited numbers of base
  station antennas,'' \emph{{IEEE} Trans. Wireless Commun.}, vol.~9, no.~11,
  pp. 3590--3600, Nov. 2010.

\bibitem{Zakhour2012}
R.~Zakhour and S.~Hanly, ``Base station cooperation on the downlink: Large
  system analysis,'' \emph{{IEEE} Trans. Inf. Theory}, vol.~58, no.~4, pp.
  2079--2106, Apr. 2012.

\bibitem{Rusek2013}
F.~Rusek, D.~Persson, B.~K. Lau, E.~Larsson, T.~Marzetta, O.~Edfors, and
  F.~Tufvesson, ``Scaling up {MIMO}: Opportunities and challenges with very
  large arrays,'' \emph{{IEEE} Signal Process. Mag.}, vol.~30, no.~1, pp.
  40--60, Jan. 2013.

\bibitem{JSAC_Large_Scale}
\emph{Special Issue: Large-scale Multiple Antenna Wireless Systems, {IEEE} J.
  Sel. Areas Commun.}, vol.~31, no.~2, Feb. 2013.

\bibitem{Telatar1999}
E.~Telatar, ``Capacity of multi-antenna gaussian channels,'' \emph{Euro. Trans.
  Telecommun.}, vol.~10, no.~6, pp. 585--595, Nov. 1999.

\bibitem{Chizhik2002}
D.~Chizhik, G.~Foschini, M.~Gans, and R.~Valenzuela, ``Keyholes, correlations,
  and capacities of multielement transmit and receive antennas,'' \emph{{IEEE}
  Trans. Wireless Commun.}, vol.~1, no.~2, pp. 361--368, Apr. 2002.

\bibitem{Roh2002}
W.~Roh and A.~Paulraj, ``{MIMO} channel capacity for the distributed antenna,''
  in \emph{Proc. IEEE VTC}, vol.~2, Sep. 2002, pp. 706--709.

\bibitem{Zhuang2003}
H.~Zhuang, L.~Dai, L.~Xiao, and Y.~Yao, ``Spectral efficiency of distributed
  antenna system with random antenna layout,'' \emph{Electronics Letters},
  vol.~39, no.~6, pp. 495--496, Mar. 2003.

\bibitem{Zhang2004}
H.~Zhang and H.~Dai, ``On the capacity of distributed {MIMO} systems,'' in
  \emph{Proc. IEEE CISS}, Mar. 2004, pp. 1--5.

\bibitem{Tulino2004}
A.~Tulino and S.~Verdu, \emph{Random Matrix Theory and Wireless
  Communications}.\hskip 1em plus 0.5em minus 0.4em\relax Now Publisher Inc.,
  2004.

\bibitem{Couillet2011}
R.~Couillet and M.~Debbah, \emph{\BIBforeignlanguage{English}{Random Matrix
  Methods for Wireless Communications}}.\hskip 1em plus 0.5em minus 0.4em\relax
  Cambridge University Press, 2011.

\bibitem{Lozano2002}
A.~Lozano and A.~Tulino, ``Capacity of multiple-transmit multiple-receive
  antenna architectures,'' \emph{{IEEE} Trans. Inf. Theory}, vol.~48, no.~12,
  pp. 3117--3128, Dec. 2002.

\bibitem{Tulino2005}
A.~Tulino, A.~Lozano, and S.~Verdu, ``Impact of antenna correlation on the
  capacity of multiantenna channels,'' \emph{{IEEE} Trans. Inf. Theory},
  vol.~51, no.~7, pp. 2491--2509, Jul. 2005.

\bibitem{Aktas2006}
D.~Aktas, M.~Bacha, J.~Evans, and S.~Hanly, ``Scaling results on the sum
  capacity of cellular networks with {{MIMO}} links,'' \emph{{IEEE} Trans. Inf.
  Theory}, vol.~52, no.~7, pp. 3264--3274, Jul. 2006.

\bibitem{Feng2009b}
W.~Feng, Y.~Li, S.~Zhou, J.~Wang, and M.~Xia, ``Downlink capacity of
  distributed antenna systems in a multi-cell environment,'' in \emph{Proc.
  IEEE WCNC}, Apr. 2009, pp. 1--5.

\bibitem{Heliot2011}
F.~Heliot, R.~Hoshyar, and R.~Tafazolli, ``An accurate closed-form
  approximation of the distributed {{MIMO}} outage probability,'' \emph{{IEEE}
  Trans. Wireless Commun.}, vol.~10, no.~1, pp. 5--11, Jan. 2011.

\bibitem{Zhang2013b}
J.~Zhang, C.-K. Wen, S.~Jin, X.~Gao, and K.-K. Wong, ``On capacity of
  large-scale {MIMO} multiple access channels with distributed sets of
  correlated antennas,'' \emph{{IEEE} J. Sel. Areas Commun.}, vol.~31, no.~2,
  pp. 133--148, Feb. 2013.

\bibitem{Choi2007}
W.~Choi and J.~Andrews, ``Downlink performance and capacity of distributed
  antenna systems in a multicell environment,'' \emph{{IEEE} Trans. Wireless
  Commun.}, vol.~6, no.~1, pp. 69--73, Jan. 2007.

\bibitem{Lee2012}
S.-R. Lee, S.-H. Moon, J.-S. Kim, and I.~Lee, ``Capacity analysis of
  distributed antenna systems in a composite fading channel,'' \emph{{IEEE}
  Trans. Wireless Commun.}, vol.~11, no.~3, pp. 1076--1086, Mar. 2012.

\bibitem{Wang2013}
D.~Wang, J.~Wang, X.~You, Y.~Wang, M.~Chen, and X.~Hou, ``Spectral efficiency
  of distributed {MIMO} systems,'' \emph{{IEEE} J. Sel. Areas Commun.},
  vol.~31, no.~10, pp. 2112--2127, Oct. 2013.

\bibitem{Gesbert2007}
D.~Gesbert, M.~Kountouris, R.~Heath, C.-B. Chae, and T.~Salzer, ``From single
  user to multiuser communications: Shifting the {MIMO} paradigm,''
  \emph{{IEEE} Signal Process. Mag.}, vol.~24, no.~5, pp. 36--46, Sep. 2007.

\bibitem{Spencer2004}
Q.~Spencer, A.~Swindlehurst, and M.~Haardt, ``Zero-forcing methods for downlink
  spatial multiplexing in multiuser {{MIMO}} channels,'' \emph{{IEEE} Trans.
  Signal Process.}, vol.~52, no.~2, pp. 461--471, Feb. 2004.

\bibitem{Shen2006}
Z.~Shen, R.~Chen, J.~Andrews, R.~Heath, and B.~Evans, ``Low complexity user
  selection algorithms for multiuser {MIMO} systems with block
  diagonalization,'' \emph{{IEEE} Trans. Signal Process.}, vol.~54, no.~9, pp.
  3658--3663, Sep. 2006.

\bibitem{Shen2007}
------, ``Sum capacity of multiuser {MIMO} broadcast channels with block
  diagonalization,'' \emph{{IEEE} Trans. Wireless Commun.}, vol.~6, no.~6, pp.
  2040--2045, Jun. 2007.

\bibitem{Shim2008}
S.~Shim, J.~S. Kwak, R.~Heath, and J.~Andrews, ``Block diagonalization for
  multi-user {MIMO} with other-cell interference,'' \emph{{IEEE} Trans.
  Wireless Commun.}, vol.~7, no.~7, pp. 2671--2681, Jul. 2008.

\bibitem{Ravindran2008}
N.~Ravindran and N.~Jindal, ``Limited feedback-based block diagonalization for
  the {MIMO} broadcast channel,'' \emph{{IEEE} J. Sel. Areas Commun.}, vol.~26,
  no.~8, pp. 1473--1482, Aug. 2008.

\bibitem{Liu2012}
Z.~Liu and L.~Dai, ``Asymptotic per-user rate analysis of downlink {{MIMO}}
  cellular networks with linear precoding,'' \emph{{IEEE} Trans. Wireless
  Commun.}, vol.~11, no.~12, pp. 4536--4548, Dec. 2012.

\bibitem{Li2009}
X.~Li, M.~Luo, M.~Zhao, L.~Huang, and Y.~Yao, ``Downlink performance and
  capacity of distributed antenna system in multi-user scenario,'' in
  \emph{Proc. IEEE WiCOM}, Sep. 2009, pp. 1--4.

\bibitem{Wang2009a}
T.~Wang, Y.~Wang, K.~Sun, and Z.~Chen, ``On the performance of downlink
  transmission for distributed antenna systems with multi-antenna arrays,'' in
  \emph{Proc. IEEE VTC}, Sep. 2009, pp. 1 --5.

\bibitem{Ahmad2011}
T.~Ahmad, S.~Al-Ahmadi, H.~Yanikomeroglu, and G.~Boudreau, ``Downlink linear
  transmission schemes in a single-cell distributed antenna system with port
  selection,'' in \emph{Proc. IEEE VTC}, May 2011, pp. 1--5.

\bibitem{Heath2011}
R.~Heath, T.~Wu, Y.~H. Kwon, and A.~Soong, ``Multiuser {{MIMO}} in distributed
  antenna systems with out-of-cell interference,'' \emph{{IEEE} Trans. Signal
  Process.}, vol.~59, no.~10, pp. 4885--4899, Oct. 2011.

\bibitem{Dai2011}
L.~Dai, ``A comparative study on uplink sum capacity with co-located and
  distributed antennas,'' \emph{{IEEE} J. Sel. Areas Commun.}, vol.~29, no.~6,
  pp. 1200 --1213, Jun. 2011.

\bibitem{Marcenko1967}
V.~A. Marcenko and L.~A. Pastur, ``Distribution of eigenvalues for some sets of
  random matrices,'' \emph{Mathematics of the USSR-Sbornik}, vol.~1, no.~4, pp.
  457--483, 1967.

\bibitem{Marcus1964}
M.~Marcus and H.~Minc, \emph{Survey of Matrix Theory and Matrix
  Inequalities}.\hskip 1em plus 0.5em minus 0.4em\relax Allyn and Bacon, Inc.,
  1964.

\end{thebibliography}

\end{document}